\documentclass[]{aa}

\usepackage{graphicx,color}
\usepackage{natbib}
\usepackage{txfonts}
\newcommand      \qj     {J1148+5251}
\newcommand   \lbg    {MS 1512-cB58}
\newcommand   \et     {et~al.}
\begin{document}

\title{The chemical evolution of elliptical galaxies with stellar and 
QSO dust production }

%\author{A. Pipino\inst{1,2} \thanks {email to: pipino@oats.inaf.it} \and  X.L. Fan \inst{2,3}   \and F. Matteucci\inst{2,3} 
%\and F. Calura\inst{3,4} \and L.Silva\inst{3}  \and G. Granato\inst{3}\and R.Maiolino\inst{5}} \institute{
% Department of Physics and Astronomy, University of
% California Los Angeles, Los Angeles CA 90025, U.S.A \and Dipartimento di Fisica, sez.Astronomia, Universit\`a di Trieste, via
% G.B. Tiepolo 11, I-34131, Trieste, Italy \and I.N.A.F. Osservatorio Astronomico di
% Trieste, via G.B. Tiepolo 11, I-34131, Trieste, Italy \and
%Jeremiah Horrocks Institute for Astrophysics and Supercomputing, University of Central Lancashire, Preston PR1 2HE \and
%Osservatorio Astronomico di Roma, INAF, Via di Frascati 44, 00040 Monte Porzio Catone, Italy \and}

\author{A. Pipino\inst{1,5} \thanks {email to: pipino@oats.inaf.it} \and  X.L. Fan \inst{1,2}   \and F. Matteucci\inst{1,2} 
\and F. Calura\inst{2,3} \and L.Silva\inst{2}  \and G. Granato\inst{2}\and R.Maiolino\inst{4}} \institute{
 Dipartimento di Fisica, sez.Astronomia, Universit\`a di Trieste, via
 G.B. Tiepolo 11, I-34131, Trieste, Italy \and I.N.A.F. Osservatorio Astronomico di
 Trieste, via G.B. Tiepolo 11, I-34131, Trieste, Italy \and
Jeremiah Horrocks Institute for Astrophysics and Supercomputing, University of Central Lancashire, Preston PR1 2HE \and
Osservatorio Astronomico di Roma, INAF, Via di Frascati 44, 00040 Monte Porzio Catone, Italy \and
 Department of Physics and Astronomy, University of
 California Los Angeles, Los Angeles CA 90025, U.S.A }

\titlerunning{The Chemical Evolution Of Elliptical Galaxies With Dust }

\authorrunning{Pipino et al.}

\date{Received xxxx / Accepted xxxx}

\abstract%{context}{aims}{methods}{results}{conclusions}
{The presence of dust strongly affects the way we see galaxies and also the chemical abundances we measure in gas. It is therefore important to study the chemical evolution of galaxies by taking into account dust evolution.}
{We aim at performing a detailed study of abundance ratios of high redshift objects and their dust properties. We focus on Lyman-Break galaxies (LBGs) and Quasar (QSO) hosts and likely progenitors of low- and high-mass present-day elliptical galaxies, respectively.} 
{We have adopted a chemical evolution model for elliptical galaxies taking account the dust production from low and intermediate mass stars, supernovae Ia, supernovae II, QSOs and both dust destruction and accretion processes. By means of such a model we have followed the chemical evolution of ellipticals of different baryonic masses. Our model complies with \emph{chemical downsizing}.}
{We made predictions for the abundance ratios versus metallicity trends for models of differing masses that
can be used to constrain the star formation rate, initial mass function and dust mass in observed galaxies.
We predict the existence of a high redshift dust mass-stellar mass relationship.
We have found a good agreement with the properties of LBGs if we assume that they formed at redshift z=2-4. 
In particular, a non-negligible amount of dust is needed to explain the observed abundance pattern.
We studied the QSO SDSS J114816, one of the most distant QSO ever observed (z=6.4), and we have been able to reproduce the amount of dust measured in this object. The dust is clearly due to the production from supernovae and the most massive AGB stars as well as from the
grain growth in the interstellar medium. The QSO dust is likely to dominate only in the very central regions
of the galaxies and during the early development of the galactic wind. }{}

\keywords{ISM: dust, extinction; ISM: abundances; galaxies: ellipticals: chemical abundances, formation and evolution ; individual: SDSS~J114816.64+525150.3, MS1512-cB58.}

\maketitle
\section{Introduction}
Dust modifies the physical and chemical conditions of the interstellar medium (ISM), in particular it
prevents stellar radiation from penetrating dense clouds, affects the cooling process, and  depletes elements into grains. Dust absorbs and scatters photons mostly at wavelengths $\lesssim 1 \mu m $, and  reemits them as infrared photons. Those processes affect our interpretation of the observational results, for instance those concerning the spectral energy distribution (SED) and the chemical abundances of the gas. Dust is a fundamental component which cannot be neglected especially when comparing
model results to high redshift observations.
There are clear evidences that a non-negligible amount of dust exists at high redshift as indicated by the
extinction curves for Lyman break galaxies (LBGs) \citep{shapley01_lbg}
%and by the depletion of heavy elements in quasar absorption systems {(ref??)}, 
and by the far infrared (IR) to millimeter (mm) dust thermal emission of distant quasars \citep{Bellen06_dust_quasar_data,bertoldi03a,wang08_dust_quasar_data}.
The main stellar sites for dust production are the cool dense atmospheres of Asymptotic Giant Branch (AGB) stars and the supernovae (SNe), and the type of dust formed depends on the chemical composition of the ejecta. Despite the fact that AGB stars and SNe Ia contribute to dust already after 30 Myr (i.e. the lifetime of a 8 $M_{\odot}$) since the beginning of star formation, the large observed amount of dust at  high redshift in some quasars needs perhaps an additional dust source.  The SNe II are one obvious stellar site (Maiolino et al. 2004), but it is not clear how dust evolution proceeds in SN II enviroments \citep{dwek2007b,Kozasa2009}.
{ In fact, while submillimetre observations of the Kepler and CasA SN remnants yield 0.1-1 $M_{\odot}$ of dust (e.g. Morgan et al., 2003,
Dunne et al., 2009, Gomez et al., 2009), infrared observations give $<<$ 0.1 $M_{\odot}$ of dust (Helmhamdi et al., 2003, 
Sugerman, et al. 2006, Kotak et al., 2009; Rho J. et al., 2008, 2009; Meikle et al., 2007; Takaya et al. 2008). A possibility is that such  a discrepancy could be traced back to a difference in the
instrument sensitivity to the different dust phases (cold for submillimetre,
warm for Spitzer). On the other hand, theoretical calculations may yield up to 1 $M_{\odot}$ of 
dust, with the exact value depending on the progenitor mass and chemical composition of the ejecta\footnote{For instance 
in Calura et al. (2008) we assume that a 20 $M_{\odot}$ star forms 0.8 $M_{\odot}$ of dust} (e.g. Dwek, 1998 and
references therein). However, a large fraction of this dust can be destroyed by the reverse shock
inside the SN remnant (e.g. Bianchi \& Schneider, 2007), with the details depending on the degree
of asymmetry in the explosion (Nozawa et al., 2010).}

The so-called smoking quasars, namely quasars produced in the clouds
in the broad line regions \citep{elvis2002}, might be one of the few sources of high redshift dust, but how much dust can be formed by the QSO wind is still unknown. Markwick-Kemper et al.(2007) determined the composition of dust in the broad absorption line (BAL) quasar PG 2112+059. They argued that the derived crystalline silicate fraction requires high temperatures such as those found in the immediate quasar environment in order to counteract the rapid destruction from cosmic rays, thus lending support to \citet{elvis2002} idea.

The measurement of metallicity and dust depletion, which are products of the (inter-)stellar evolution, provides an important constraint on chemical models. The formation and evolution of dust in primordial SNe as well as the dust evolution effects on SED of galaxies have been studied by \cite{nozawa08} and \cite{schurer09_dust_sed}, respectively.  Studies of the gas, metal and dust evolution  have been performed by means of analytical models \citep{dwek07} and simulations \citep{dwek98,morgan03,inoue03,valiante09}. \cite{calura08dust} applied a detailed formulation of dust production and destruction to a detailed chemical evolution model.
In that paper the model was mostly constrained by means of the solar neighbourhood dust determination,  and then it was proven to be
quite flexible, since it can be readily applied to models for elliptical and irregular galaxies. The qualitative dust mass
evolution with time in high mass spheroids was depicted and Calura et al. (2008) showed how the dust is an important factor also in the very late phases of their lives, namely when the star formation is over.
In this paper we aim at studying in detail the dust evolution in elliptical galaxies by starting from the model by \cite{calura08dust}. 
We update the dust production by changing the dust condensation efficiency in supernova Ia and II (in order to take into
account the above mentioned new observational constraints)
and adding the possibility of QSO dust. We also revise the accretion process timescale. %in light of the recent results by Zhukovska et al. (2008).
We present models of ellipticals of different baryonic masses in the range $10^{9} - 10^{12}M_{\odot}$, and compare them with LBGs and with QSO hosts.
In particular, after some general predictions, we focus
on the particular cases: LBG ~  \lbg (as a proxy for a relatively low mass elliptical) and the  host galaxy of QSO ~ \qj (as a proxy
for the most massive spheroids).

The paper is organized as follows. The chemical evolution model and the dust model  are described in \S 2 and \S 3, respectively. Observational data are briefly summarized in \S 4 . We predict the chemical and  dust evolution of elliptical galaxies in different conditions and apply them to interpret the observational data in \S 5-7. Our conclusions are summarized in \S 8.

\section{The chemical evolution model for elliptical galaxy}
\subsection{The chemical evolution code}
The model adopted here is basically the multi-zone model  of Pipino \& Matteucci (2004).
The model galaxy is divided into spherical shells 0.1 $R_{eff}$ thick. Most of the results
presented below refer to global galactic properties and
are calculated by considering all the shells. { In practice, quantities like the total mass
in stars are computed by summing up the single shell contributions, whereas the chemical abundances
are (gas) mass-weighted averages. }
The galaxy evolves as  an open box in which the initial gas mass, with primordial chemical composition,
rapidly  collapses, on a time scale $\tau$,
into the potential well of a dark matter halo.

The rapid collapse triggers an intense and rapid  star formation process, which can be considered as a
starburst and lasts until a galactic  wind,
powered by the  thermal energy injected by stellar winds and SN (Ia,
II) explosions, occurs. At that time, the gas thermal energy  equates the
gas binding energy of and  all the residual interstellar medium is assumed to be
lost. After that time, the galaxies evolve passively.

The equation of chemical evolution for the element $i$ in each galactic shell takes the following form:
\begin{eqnarray}\label{main}
{d G_i (t) \over d t}  &= & -\psi (t) X_i (t)\,   \nonumber \\
& & +\int_{M_L}^{M_{B_m}} \psi (t-\tau_m) Q_{\rm mi}(t-\tau_m) \phi (m) dm\,  \nonumber \\
& & + A\int_{M_{B_m}}^{M_{B_M}} \phi (m) \nonumber  \\
& &\cdot \left [ \int_{\rm\mu_{\rm min}}^{0.5} f(\mu) Q_{\rm mi}(t-\tau_{\rm m_2}) \psi (t-\tau_{\rm m_2}) d\mu \right ] dm \,  \nonumber \\
& &  +(1-A)\, \int_{\rm M_{\rm B_m}}^{M_{\rm B_M}} \psi (t-\tau_m) Q_{\rm mi}(t-\tau_m) \phi (m) dm\,  \nonumber \\
& &  +\int_{\rm M_{\rm B_M}}^{M_U} \psi (t-\tau_m) Q_{\rm mi}(t-\tau_m) \phi (m) dm\,  \nonumber \\
& & +({d G_i (t) \over d t})_{\rm infall}
%-W(t)X_i (t)
%+({d G_i (t) \over d t})_{\rm acc}
\, ,
\end{eqnarray}
where $G_i (t)= M_{gas}(t) \,X_i (t) $
is the  fractional mass  of the element \emph{i} at the time \emph{t} in the ISM.
The quantity $X_i (t)$ is defined as the abundance by mass of the element 
\emph{i}.
By definition $\sum_i X_i=1$.

The first term  on the right side of eq. \ref{main} gives the rate at which the element
$i$ is  subtracted from ISM by the SF process.  The variable $\psi$ is
the star formation rate calculated according to the following law:
\begin{equation}
\psi (t)= \nu\cdot M_{gas} (t) \, ,
\end{equation}
namely it is assumed to be proportional to the gas mass
via a constant $\nu$ which represents the star formation efficiency. In order to reproduce the 'inverse wind
model' of  \citep{m94_inv_wind}, an earlier version of the now popular "downsizing",
we assume $\nu$ as an increasing function of the galactic mass,
(see Table \ref{t1}) { following the findings of our previous work (Pipino \& Matteucci, 2004).}

The second term is the rate at which each element is restored into
the ISM by single stars with masses in the range $M_{L}$ -
$M_{B_m}$, where $M_{L}$ is the minimum mass contributing, at a
given time $t$, to chemical enrichment  and $M_{B_m}$ is the minimum binary mass allowed for
binary systems giving rise to SNIa ($3 M_{\odot}$, Greggio \& Renzini, 1983).
The initial mass function (IMF) is $\phi (m)\propto m^{-(1+1.35)}$, (Salpeter 1955), and it is
normalized to unity in the mass interval $0.1 -100 M_{\odot}$. In
particular, $Q_{mi}(t-\tau_m)$,
is a matrix which calculates for any star of a given mass $m$ the
amount of the newly processed and already present element $i$, which
is returned to the ISM. The quantity $\tau_m$ is the lifetime of a
star of mass $m$ (Padovani \& Matteucci, 1993).

The third term represents the enrichment by SN Ia for which we assume
the single degenerate scenario: a C-O white dwarf plus a red giant (Whelan \& Iben 1973).  
We refer to Greggio $\&$ Renzini (1983), Matteucci $\&$ Greggio (1986) and Matteucci \& Recchi (2001)
for further details.
%, the formalism is:
%\begin{equation}
%R_{SNIa}=A\int^{M_{BM}}_{M_{Bm}}{\phi(M_B) \int^{0.5}_{\mu_{min}}{f(\mu)
%\psi(t- \tau_{M_{2}})d \mu \, dM_{B}}}\, ,
%\end{equation}
%where $M_{\rm B}$ is the total mass of the
%binary system and $M_{BM}=16 M_{\odot}$ is the
%maximum mass allowed for the adopted progenitor systems.
%$\mu=M_2/M_{\rm B}$ is the mass fraction of the secondary, which
%is assumed to follow the distribution law:
%\begin{equation}
%f(\mu)=2^{1+\gamma}(1+\gamma)\mu^\gamma\, .
%\end{equation}
%Finally, $\mu_{min}$ is the minimum mass fraction contributing to the SNIa rate at the time $t$, and is given by:\\
%\begin{equation}
%\mu_{min}=max \left \{ \frac{M_{2}(t)}{M_{B}}, \frac{M_{2}-0.5M_{B}}{M_{B}} \right \}
%\end{equation} and $\gamma=2$.
The predicted SNIa explosion rate is constrained
to reproduce the present day observed value (Mannucci et al., 2008), by
fixing the parameter $A=0.09$ in eq. (\ref{main}). 
%In particular, $A$ represents
%the fraction of binary systems in the IMF which are
%able to give rise to SNe Ia explosions. In the following we adopt $A=0.09$.

The fourth and fifth terms represent the enrichment by single massive stars. 
%, the SNII rate is:
%\begin{eqnarray}
%R_{SNII} & = & (1-A)\int^{16}_{8}{\psi(t-\tau_m) \phi(m)dm}\nonumber \\
%& + & \int^{M_U}_{16}{\psi(t-\tau_m) \phi(m)dm}\, ,
%\end{eqnarray}
%where the first integral accounts for the single stars in the
%range 8-16$M_{\odot}$, and $M_{U}$
%is the upper mass limit in the IMF.
The initial galactic infall phase enters the equation via the sixth
term, for which we adopt the formula:
\begin{equation}
({d G_i (t) \over d t})_{\rm infall}= X_{i,\rm infall} C e^{-{t \over
\tau}}\, ,
\end{equation}
where $X_{i,\rm infall}$ describes the chemical composition of the
accreted gas, assumed to be primordial. $C$ is the normalization
constant obtained by integrating the infall law over a Hubble time. { For instance,
for a $10^{11} M_{\odot}$ model, the accretion history is such that \rm
90\% of the initial gas has already been accreted at $t_{gw}$ (in fact, we
halt the infall of the gas at the occurrence of the galactic wind).}
Finally, in order to calculate the potential energy of gas, we take into account the presence of a dark matter halo.
The assumed prescriptions are the same as in our previous models
(e.g. Pipino \& Matteucci 2004).

\subsection{Stellar yields}

The yields used in this paper are as follows:
\begin{enumerate}
\item
For single low and intermediate mass stars ($0.8 \le M/M_{\odot} \le
8$) we make use of the yields by van den Hoek \& Gronewegen (1997)  as a function of metallicity.
\item For SNIa and SNII we adopt the recently suggested empirical yields
by Fran\c cois et al. (2004). These yields are a revised version of the Woosley
\& Weaver (1995, for SNII) and Iwamoto et al. (1999, for SNIa) calculations adjusted
to best fit the chemical abundances in the Milky Way.
When discussing \qj we will also present a case in which for O and C we have adopted
the yields computed assuming mass loss in massive stars for $Z > Z_{\odot}$ as in Maeder (1992).
In fact, including mass loss in massive stars produces a large loss of C and He, thus 
lowering the O production, and this effect is significant only for over-solar metallicity. 
Recent papers (McWilliam \& al. 2008; Cescutti \& al. 2009) have shown that these yields are required 
to reproduce the [O/Mg] and [C/O] ratios at high metallicities in the Galactic bulge.
\end{enumerate}

\begin{table*}
%\centering
\begin{minipage}{120mm}
\scriptsize
\begin{flushleft}
\caption{Model parameters and results.}\label{t1}
\begin{tabular}{l|llllll|ll|lllllll}
\hline
\hline
model          &                &           &              &                  &             &             &           &      & &                       \\
name           & $M_{lum}$ 	&$R_{eff}$  &  $\nu$ 	   & $\tau$& $t_{gw}$ &$[<Mg/Fe>_*]$& C,O yields &        N & stellar& recipe & $\tau_0$ & QSO &$M_{BH,seed}$ & $\tau_{BH}$ & BEL\\
               &({$M_{\odot}$}) & ({kpc}) &  ({$Gyr^{-1}$})& (Gyr)& (Gyr)     &             &            & primary     & dust   &               &         & dust&({$M_{\odot}$})& (Gyr) & metall.\\
\hline
M9             &$10^{9}$       & 0.4        & 1.2            & 0.8    &  1.2    & +0.28	    & Fran\c cois&    no     &  off  & - & - & - & - & - & - \\
M9d            &$10^{9}$       & 0.4        & 1.2            & 0.8    &  1.2    & +0.28	    & Fran\c cois&    no     &  on   & this work    &  0.1    & off & - & - & -\\
M10            &$10^{10}$       & 1         & 3              & 0.5    &  1.1    & +0.29	    & Fran\c cois&    no     &  off  & - & - & - & - & - & -\\
M10d           &$10^{10}$       & 1         & 3              & 0.5    &  1.1    & +0.29	    & Fran\c cois&    no     &  on   & this work    &  0.08    & off & - & - & -\\
M310           &$3\cdot 10^{10}$& 2         & 5              & 0.5    &  1.0    & +0.30	    & Fran\c cois&    no     &  off  & - & - & - & -& - & -\\
M310N          &$3\cdot 10^{10}$& 2         & 5              & 0.5    &  1.0    & +0.30	    & Fran\c cois&    yes    &  off  & - & - & - & - & -\\
M310d          &$3\cdot 10^{10}$& 2         & 5              & 0.5    &  1.0    & +0.30	    & Fran\c cois&    no     &  on   & this work    &  0.05    & off & - & - & -\\
M310Nd         &$3\cdot 10^{10}$& 2         & 5              & 0.5    &  1.0    & +0.30	    & Fran\c cois&    yes    &  on   & this work    &  0.05    & off & - & - & -\\
M310Cd         &$3\cdot 10^{10}$& 2         & 5              & 0.5    &  1.0    & +0.30	    & Fran\c cois&    no     &  on   & Calura       &  0.05    & off & - & - & -\\
M310MD         &$3\cdot 10^{10}$& 2         & 5              & 0.5    &  1.0    & +0.30	    & Maeder     &    no     &  off  & - & - & - & -& - & -\\
M310MDd        &$3\cdot 10^{10}$& 2         & 5              & 0.5    &  1.0    & +0.30	    & Maeder     &    no     &  on   & this work    &  0.01    & off & - & - & -\\
M11            &$10^{11}$       & 3         & 10             & 0.4    &  0.9    & +0.33     & Fran\c cois&    no     &  off  & - & - & - & - & - & -\\
M11d           &$10^{11}$       & 3         & 10             & 0.4    &  0.9    & +0.33	    & Fran\c cois&    no     &  on   & this work    &  0.03    & on & - & - & -\\
M12            &$10^{12}$       & 10        & 20             & 0.2    &  0.7    & +0.39     & Fran\c cois&    no     &  off  & - & - & - & - & - & -\\
M12d           &$10^{12}$       & 10        & 20             & 0.2    &  0.7    & +0.39	    & Fran\c cois&    no     &  on   & this work    &  0.01    & on &$2\cdot 10^3$ & 0.049 & mod\\
M12MDd         &$10^{12}$       & 10        & 20             & 0.2    &  0.7    & +0.39     & Maeder     &    no     &  on   & this work    &  0.01    & on &$2\cdot 10^3$ & 0.049 & mod\\ 
M12dcaseA      &$10^{12}$       & 10        & 20             & 0.2    &  0.7    & +0.39	    & Fran\c cois&    no     &  on   & this work    &  0.01    & on &$2\cdot 10^3$ & 0.049 & 5x\\
M12dcaseB      &$10^{12}$       & 10        & 20             & 0.2    &  0.7    & +0.39	    & Fran\c cois&    no     &  on   & this work    &  0.01    & on &$4\cdot 10^3$ & 0.049 & mod\\
M12dcaseC      &$10^{12}$       & 10        & 20             & 0.2    &  0.7    & +0.39	    & Fran\c cois&    no     &  on   & this work    &  0.01    & on &$8\cdot 10^3$ & 0.049 & mod\\
M12dcaseD      &$10^{12}$       & 10        & 20             & 0.2    &  0.7    & +0.39	    & Fran\c cois&    no     &  off  &  -           &  0.01    & on &$1\cdot 10^8$ & 0.49 & mod\\

\hline
\end{tabular}
\end{flushleft}
The values for $t_{gw}$ and  $[<Mg/Fe>]$ refer to the central regions of the galaxy. These values are determined only by $M_{lum}$ (and hence $\nu$ and $\tau$).
At each given mass models differ by C, O and N yields - fiducial, Maeder (1992) and primary production, respectively - dust (on/off, prescriptions
for stellar dust, $\tau_0$ and QSO dust production). The BEL metallicity is the metallicity of the gas out of which the QSO dust forms: \emph{mod} 
refer to a model self-consistently calculated metallicity, \emph{5x} is for the case in which the adopted metallicity is 5 times higher than the average \emph{mod} one.
\end{minipage}
\end{table*}

%\subsection{Energetics}
%The two competing forms of energy that set the galatic wind time are: i) the thermal energy of
%gas (including the stellar feedback and gas cooling) and ii) the potential energy of the gas.
%We assume that stellar winds and  SNe Ia, and II are the heating sources.
%The supernova II includes all core-collapse SNe and we consider that they all produce the same amount of energy ($\sim 10^{51}$ erg).
%We assume that a galactic wind occurs when the gas thermal energy is larger than the gas potential energy. 
%When the above condition is reached, the gas flows out of the galaxy.
%The potential energy of gas takes into account the presence of a dark matter halo.
%The assumed prescriptions are the same as in our previous models
%(e.g. Pipino \& Matteucci 2004). %which were successful in reproducing the main features of ellipticals.
%In particular, we assume a dark matter halo 10 times more massive than the luminous mass but much more extended (see Matteucci 1994; Pipino \& Matteucci, 2004).

\section{Dust model}

\subsection{Stellar dust}

We adopt the  dust model of \cite{calura08dust}  which uses the formalism developed by \cite{dwek98}.  Let us define $X_{dust,i}(t)$ as the abundance by mass of the element $i$ at the time $t$ in the dust and since $G(t)$ is the ISM fraction at the time $t$, the quantity $G_{dust,i}=X_{dust,i}\cdot G(t)$ represents the normalized mass of the element $i$ at the time $t$ in the dust.
The time evolution of $G_{dust,i}$ is therefore computed as:

\begin{eqnarray}\label{eq_dust}
{d G_{dust,i} (t) \over d t} & = &  -\psi(t)X_{dust,i}(t)\nonumber\\
& & + \int_{M_{L}}^{M_{B_m}}\psi(t-\tau_m) \delta^{SW}_{i}
Q_{mi}(t-\tau_m)\phi(m)dm \nonumber\\
& & + A\int_{M_{B_m}}^{M_{B_M}}
\phi(m)\nonumber \\
& & \cdot[\int_{\mu_{min}}
^{0.5}f(\mu)\psi(t-\tau_{m2}) \delta^{Ia}_{i}
Q_{mi}(t-\tau_{m2})d\mu]dm\nonumber \\
& & + (1-A)\int_{M_{B_m}}^
{M_{w}}\psi(t-\tau_{m})  \delta^{SW}_{i} Q_{mi}(t-\tau_m)\phi(m)dm\nonumber \\
& & + (1-A)\int_{M_{w}}^
{M_{B_M}}\psi(t-\tau_{m})  \delta^{II}_{i} Q_{mi}(t-\tau_m)\phi(m)dm\nonumber \\
& & + \int_{M_{B_M}}^{M_U}\psi(t-\tau_m)  \delta^{II}_{i} Q_{mi}(t-\tau_m)
\phi(m)dm \nonumber\\
& & - \frac{G_{dust,i}}{\tau_{destr}} + \frac{G_{dust,i}}{\tau_{accr,i}} + \delta^{qso}_{i}X_{i} \psi_{f}(t)  %-({d G_{dust,i} (t) \over dt})_{\rm out} \\
\end{eqnarray}
where $M_w=8M_{\odot}$.
{ We refer the reader to Calura et al. (2008)  for further details and we note that in this equation we neglected 
the \emph{wind} term which allows us to take into account the fraction of dust which is ejected during the galactic wind.
The main reason is that we are focusing on the $t<t_{gw}$ evolution. The dust mass after the wind depends on several assumptions (e.g. fate of the dust
in the wind stage, wind mass loading, sputtering in the hot medium) and we are planning to improve upon the Calura et al. (2008) formulation
 in a future work.}
Here recall the general features of the model as
well as the main assumptions. Only the main refractory elements, C, O, Mg, Si, S, Ca, Fe, are depleted into dust,
and we assume that stars can produce two different types of grains: i) silicate dust, composed of O, Mg, Si, S, Ca and Fe, and ii) carbon dust, composed of C.  As  suggested by \cite{dwek98}, we consider that the dust producers are  low and intermediate mass stars,  SNIa and  SNII.
The condensation efficiencies $\delta_i^{SW}$, $\delta_i^{Ia}$ and $\delta_i^{II}$, for low and intermediate mass stars, SNIa and SNII, respectively, are as follows.

{ In low and intermediate-mass stars, dust is produced during the Asymptotic Giant Branch (AGB) phase 
(Ferrarotti \& Gail 2006 and references therein).  We assume that 
dust formation depends mainly on the composition of the stellar envelopes. 
{If $X_{O}$ and $X_{C}$ represent the O and C mass fractions in the stellar envelopes, respectively, we assume that } 
stars with $X_{O}/X_{C}$ $>$ 1 are producers of silicate dust, i.e. dust particles composed by O, Mg, Si, S, Ca, Fe. On the other hand, C rich stars, 
characterized by $X_{O}/X_{C}$ $<$ 1, are producers of carbonaceous solids, i.e. carbon dust (Draine 1990). 
Being { $M_{ej, i}(m) \propto m\times Q_{\rm mi}(t) $ and $M_{d,i}(m)$} the total ejected mass and the dust mass formed by a star of initial mass $m$ 
for the element $i$, 
respectively, we assume that for stars with $X_{O}/X_{C}$ $<$ 1:}
\\
\\
\begin{math}
M_{d, C}(m) =  \delta^{SW}_{C} \cdot [M_{ej,C}(m)-0.75 M_{ej,O}(m)]
\end{math}
with $\delta^{SW}_{C}=1$ 
\\
\\
\begin{math}
M_{d, i}(m) =0, 
\end{math}
for all the other elements. \\
\\
{ For stars with $X_{O}/X_{C}$ $>$ 1 in the envelope, we assume:} \\
\\
\begin{math}
M_{d, C}(m) =0  \, ,M_{d, i}(m) = \delta^{SW}_{i} M_{ej, i}(m) \propto \delta^{SW}_{i} Q_{\rm mi}\\
\end{math}
\\
with $\delta^{SW}_{i}=1$ for Mg, Si, S, Ca, Fe and \\
\\
\begin{math}
M_{d, O}(m)=16 \sum_{i} \delta^{SW}_{i} M_{ej, i}(m)/\mu_{i}   
\end{math}
\\
{ with $\mu_{i}$ being the mass of the $i$ element in atomic mass units. \\
For SNIa we assume:}\\
\\
\begin{math}
M_{d, i}(m) = \delta^{Ia}_{i}M_{ej, i}(m) \propto \delta^{Ia}_{i} Q_{\rm mi}\, ,
M_{d, O}(m)=16 \sum_{i} \delta^{Ia}_{i} M_{ej, i}(m)/\mu_{i}.  \\
\end{math}
\\
{ For SNII, we adopt the same prescriptions as for SNIa.}
% with the  $\delta^{Ia}_{i}$ replaced by  $\delta^{II}_{i}$. 

{ The terms $\tau_{accr}$ and $\tau_{destr}$ represent the timescales for the destruction and accretion of dust, respectively.
For a given element $i$,  
the accretion timescale $\tau_{accr}$ can be expressed as: }
\begin{equation}
\tau_{accr,i}=\tau_{0,i}/(1 - f_i) 
\label{accr_t}
\end{equation}
where 
\begin{equation}
f_i=\frac{G_{dust,i}}{G_{i}}\, .
\end{equation} 
{ Dust accretion occurs in dense molecular clouds, 
where volatile elements can condensate onto pre-existing grain cores, originating a volatile 
part called mantle (Dwek, 1998, Inoue, 2003).  Direct evidences for dust accretion come from the observed  
large variations of the depletion levels as a function of density (Savage \& Sembach 1996) and from the observed 
 infrared emission of cold molecular clouds (Flagey et al. 2006), which is characterized by the absence 
of small grain emission. 
These features can be accounted for by the coagulation of small grains on and into larger particles. 
Indirect evidence for dust accretion 
comes from the estimation of the grain lifetimes, 
which would be very small if no process could allow the grains to recondense and grow (McKee 1989, Draine \& Salpeter 1979) .}

{ Dust destruction is primarily due to the propagation 
of SN shock waves in the warm/ionized interstellar medium (McKee 1989, Jones et al. 1994). 
Following the suggestions by McKee (1989) and D98, for a given element $i$ 
the destruction timescale $\tau_{destr}$ can be expressed as: 
\begin{equation}
\tau_{destr, i}=(\epsilon M_{SNR})^{-1} \cdot \frac{\sigma_{gas}}{R_{SN}}
\label{destreq}
\end{equation} 
Hence the destruction timescale is independent of the dust mass.  
$M_{SNR}$ is the mass of the interstellar gas swept up by the SN remnant. 
For this quantity, McKee (1989) suggests 
a typical value of $ M_{SNR} \sim 6800 M_{\odot}$ as well as a
value for the destruction efficiency $\epsilon$  in a three-phase medium 
as the present-day local ISM are around 0.2, Hence Calura et al. (2008) assumed: } 
\begin{equation}
\epsilon M_{SNR} = 0.2 \times 6800 M_{\odot} = 1360 M_{\odot} \, .
\end{equation}
{ In Calura et al. (2008), $R_{SN}$ is the \emph{total} SNe rate at any given timestep, 
including the contributions by both SNIa and SNII. }

Below we discuss the different prescriptions adopted here with respect to the original model presented in Calura et al. (2008).\\
We note here that the modifications have been made through changes in the parameters (e.g. $\delta_i^{II}$, $\tau_{0,i}$),
whereas the general scheme and the above equations hold for both the Calura et al. and the present
formulations.
%or dust production (SNeII, Ia, AGB stars), destruction and accretion are the same as in Calura et al. 
%(2008) except for the following items:\\

\begin{figure}
\includegraphics[width=9cm,height=5cm]{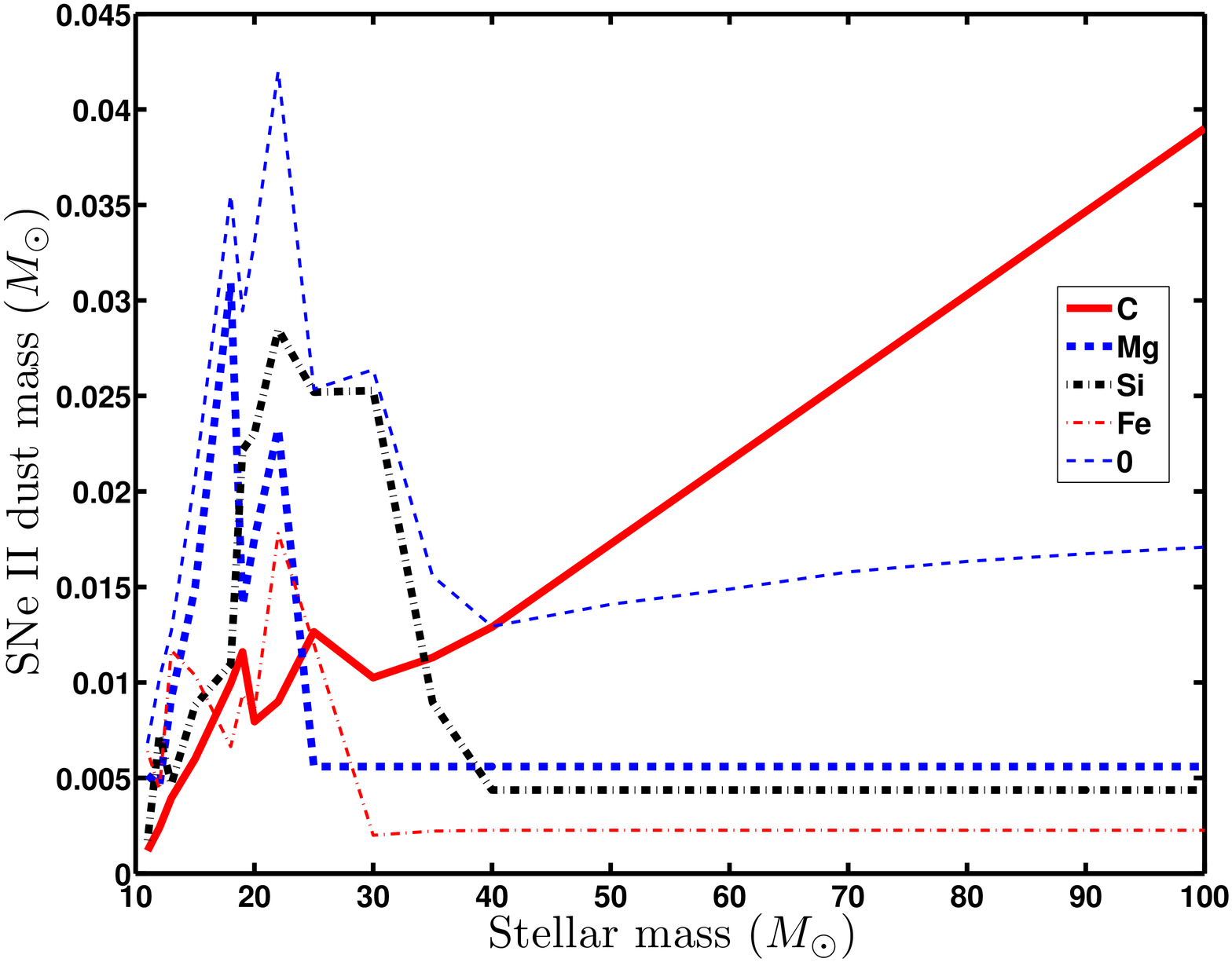}
\caption{Dust ``yields'' from SNe II as functions of initial stellar mass adopted in this paper (see description 
in the text).}
\label{snii}
\end{figure}

\emph{Dust production:}
\begin{itemize}
\item { We reduced the values for $\delta_i^{II}$ in order to
reproduce recent observational constraints (e.g. Kotak et al., 2009, Gomez et al., 2009).
In practice, since the main unknown quantity is the efficiency with which newly created dust might be destroyed 
by reverse shocks inside the SN remnant - that we cannot model - before it either becomes observable or is ejected in to the ISM,
we include this uncertainty in the quantity $\delta_i^{II}$
(see also Valiante et al., 2009a, and Zhukovska et al., 2008).
In particular, we decreased the $\delta_i^{II}$ by a factor of 10 with respect to the Calura et al.'s fiducial case.
This implies that a typical 20$M_{\odot}$ star now produces nearly 0.08$M_{\odot}$ of dust.
This is  2-3 times higher than the maximum amount of warm dust observed in SNII (Rho et al., 2008, 2009),
yet lower than the most recent estimates of the cold dust mass in the Kepler and CasA SN remnants (Gomez
et al., 2009, Dunne et al., 2009).}
The dust ``yields'' as a function of stellar mass for SNeII are shown in Fig.~\ref{snii},
{ whereas a comparison between the new efficiencies and those in Calura et al. is shown in Table~\ref{tablediff}.}
\item We reduced by a factor of 10 also the values for $\delta_i^{Ia}$ and the reason is that there is no clear indication that SNe Ia produce dust and certainly they cannot produce more dust than SNe II.
\end{itemize}

\emph{Dust accretion in the ISM:}

%\begin{itemize}
%\item 
We have improved the calculation of the dust growth rate 
for systems with a SFR (and hence a cloud destruction rate) much
higher than in the MW.
%\item 

%\end{itemize}

We assumed that the dust accretion occurs only during the starburst epoch\footnote{After the galactic wind occurs 
and star formation stops there is only dust production from long living stars and SNIa.}
in dense molecular clouds, where volatile elements can condensate onto pre-existing grain cores.
In particular, in order to calculate 
the effective growth timescale we took into account the two following facts: i) the growth timescale is in the range 10-70 Myr and
quickly decreases with the metallicity of the ISM (c.f Zhukovska
et al. 2008, their  Fig. 12), and ii) the effective growth timescale depends on the fraction of ISM in molecular clouds.
%Here we assume that  $\tau_accr \sim 1/Z$, as shown in Fig.1 where the dependence of this timescale 
%on the age is also indicated for a massive galaxy ($M=10^{12} M_{\odot}$).
As far as the former is concerned, given the high SF rates in model ellipticals (see below) the metallicity
quickly increases; therefore, for a sizeable fraction of the galaxy evolution, we are in the regime in which $\tau_0$ is shorter
than the typical survival time of molecular clouds, (20-30 Myr, e.g. Krumholz \& al. 2006). This implies that
the dust growth regime is basically determined by the cycling frequency of the ISM
between clouds rather than by the inner
processes occurring inside the clouds (Zhukovska et al., 2008), which we will neglect. 
{ Namely, if we consider a fraction of ISM which collapses into a cloud which is in turn  dispersed
by the young stars, there is always a net increase in the dust mass.
The more frequently the single regions of the ISM collapse into clouds, form stars and become
diffuse ISM again, the faster the gain in dust mass by growth.}
%This justifies
%the adopted simplistic scaling $\tau_accr \sim 1/Z$.
In other words, the only difference between galaxies of different mass/SF history might arise
%Finally, in order to calculate 
%the effective growth timescale one has to 
%multiply
%$\tau_accr$ 
from the fraction of ISM in molecular clouds at any time and by the
speed at which these clouds are (re-)created.
In the revised monolithic formation scenario, the formation of the most
massive galaxies occurs in a very intense and short burst, with
SFRs exceeding 1000 M$_{\odot}$/yr (Fig. 2). Therefore we expect that
most of the ISM is in the form of clouds for most of the time, during
the active star forming phase.
The high mass galaxy model forms twice as fast (in terms of infall timescale,
4 times if the star formation timescale $1/ \nu$ is considered) as the
low mass galaxy model. Therefore, relatively speaking, we expect that the 
high mass model will have an effective growth timescale which is a
factor of 2-4 shorter than in the low mass model.
In this latter case we adopt the fiducial value ($\tau_0$=0.05 Gyr)
from Calura et al. (2008), whereas in the former, namely the most massive galaxy, we assume $\tau_0$=0.01 Gyr.

\emph{Dust destruction:}

We accounted  for the effect of correlated SNe II, which
strongly suppresses the destruction of the interstellar dust with respect to the case
in which all SNe could be treated as isolated and random explosions. { McKee (1989), estimates that
in an association with $>$10-40 SNeII, the dust mass which  can be destroyed amounts to $\sim 1.5-2$ times
the mass of dust destroyed by a single isolated explosion, namely the mean efficiency per SNII is a
factor of 0.05-0.2 the efficiency of an isolated explosion.} To mimic this effect
we made use of an \emph{effective} SNII rate (equal to the 1/10 of the actual SNII rate) when computing
the destruction rate (Eq.~\ref{destreq}). The SNIa explosions are not correlated, therefore we still used their actual rate
for computing the destruction due to them.

Table~\ref{tablediff} shows the values adopted for the dust depletion efficiencies and
both the growth and the destrunction timescale in both the old (Dwek et al., 1998, Calura et al., 2008) and
in the revised (this paper) formulation for the dust evolution.
\begin{table*}
\centering
\begin{minipage}{120mm}
\scriptsize
\begin{flushleft}
\caption{Dust parameters in the old and the new formulation}
\begin{tabular}{l l l l}
\hline
\hline
 parameter       & Calura et al.& new value& element\\
$ \delta_i^{SW}$ & 1            & 1        & C\footnote{Refer to Sec. 3.1 for the formula for the C dust.}\\
(C/O$>$1)        & 0            & 0        & O,Mg,Si,S,Ca,Ti,Fe \\
\hline
$ \delta_i^{SW}$ & 0            & 0        & C \\
(C/O$<$1)        & 1            & 1        & Mg,Si,S,Ca,Ti,Fe \\
\hline
 $\delta_i^{Ia}$ & 0.5          & 0.05    & C \\
                 & 0.8          & 0.08     & Mg,Si,S,Ca,Ti,Fe\footnote{Refer to Sec. 3.1 for the formula for the O dust.}\\
\hline
 $\delta_i^{II}$ & 0.5          & 0.05    & C \\
                 & 0.8          & 0.08     & Mg,Si,S,Ca,Ti,Fe\\
\hline
$\tau_0$         &const with gal. mass& dependent on gal. mass\\
\hline
$\tau_{destr}$   & true SNII rate& effective\footnote{A factor of 10 lower than the true rate} SNII rate\\
                 &+true SNIa rate&+true SNIa rate     \\
\hline
QSO dust         & -            & Elvis et al. (2002) \\
\label{tablediff}
\end{tabular}
\end{flushleft}
\end{minipage}
\end{table*}

\subsection{QSO dust}\label{qsodust}
A non-stellar mechanism for dust production in the early universe was proposed by Elvis et al. (2002). 
These authors showed that the physical conditions in the clouds of the Broad Emission Line (BEL) regions which 
undergo cooling and adiabatic expansion may become similar to the conditions of AGB stellar envelopes, 
and therefore may provide sites for dust formation. The QSO wind can then enhance the pressure,
hence boosting the dust production. Elvis et al. (2002) estimate that this process starts
when the BEL clouds are about 3pc from the QSO center. Also, they estimate
that dust sublimation due to the QSO radiation should be prevented by the cloud
geometry and composition even in the most luminous QSOs.

To take into account this dust production channel we assumed a simple 
model suggested by \cite{maiolino2006} with the main difference that we use a self-consistent estimate of the 
gas metallicity in the dust formation site.
The main ingredients are the following:
\begin{enumerate}
\item[1] In order to reproduce the correlation between BH and stellar mass, 
we expect a BH mass $M_{bh} \sim 0.003M_{L}$ (Ferrarese \& Cote, 2007) at the wind time $t_{gw}$.
{ Since now on, we will refer to the above ``fiducial'' set-up,
unless stated otherwise.
In Sec. 6 we will also discuss cases in which the BH-to-stellar mass ratio is higher,
as it seems to be the case at high redshifts (e.g. Lamastra et al., 2010).}

\item[2] In order to achieve (1) it is enough to let the BH grow at the Eddington rate over the time 0-$t_{gw}$  
(Padovani \& Matteucci 1993), starting from a seed mass of $\sim 2000 M_{\odot}$.

\item[3] We assume that the mass flow rate from QSO wind is $\psi_{f}(t) =0.5\cdot 10^{-8} M_{bh}(t)$ in units of $M_{\odot}/yr$ \citep{pk2002}.

\item[4] We assume that the QSO dust creation rate scales as the mass flow rate in the QSO wind. 
The factor of proportionality being the elemental abundance $X_{i}$ in the gas which is 
calculated at each timestep. { Our model predicts average metallicities in each shell.
Therefore, the abundances in the central
region of the galaxies are lower than the ones measured in BLRs (Nagao et al., 2006b,
Juarez et al., 2009). In Sec. 6 we will show a case in which we estimate
the effect of the enhanced BLR metallicity in the creation of QSO dust.}

\item[5] We use the same depletion efficiencies as for low- and intermediate-mass stars, hence $\delta_i^{qso}=\delta_i^{sw}$.
\end{enumerate}

Interestingly, several studies (e.g. Maiolino et al. 2004) show that the extinction curve
in high redshift QSOs is similar to the one expected for a medium dominated by SN dust. 
Indeed, some SNeII (and possibly ``prompt'' SNeIa, see Mannucci et al, 2006) can provide the 
metal seeds out of which the QSO dust can condense.
The effect of the QSO dust amount and composition on the spectral
properties of high redshift spheroids can possibly constrain the above scheme and it will be the topic 
of a forthcoming paper.

{ We will study the QSO contribution to the dust only in the highest mass galaxy model. The reasons
are many: i) the observations we compare with are for high redshift very massive galaxies which host QSOs; ii) the
fraction of QSO in z$\sim$3 LBGs is lower than 3\% (Reddy et al., 2008), and iii) there
is no indication of a QSO in \lbg .}

\section{Observational Data}

{In this section we briefly summarize the observations to which we will compare our model predictions
in the rest of the paper.}

\subsection{Lyman break galaxies}
From the discovery of LBGs (see Steidel et al. 1996a,b), only a handful of objects whose
abundance pattern has been studied in great detail.
One of the first of such objects for which abundance measurements were available is MS 1512-cB58,
studied by Pettini et al. (2002). 
Owing to its  gravitationally lensed nature, MS 1512-cB58 is one of the
brightest known LBGs. This object
is at $z \sim 2.73$ and has a luminous mass of
$\sim 10^{10}M_{\odot}$, a star formation rate 
$\psi (t_{sf})\sim 40 \,\rm M_{\odot}yr^{-1}$ \citep{pettini02_cb58} and an effective radius of
$r_{L}\sim 2 \rm \, kpc$ (Seitz et al. 1998),
for a $\Omega_m= 0.3,\Omega_{\Lambda}=0.7, h=0.70$
cosmology. Pettini et al. (2002) concluded that the abundances of
O, Mg and Si are
$\sim 2/5$ of their solar values, whereas { [Fe/H]=-1.15, hence
underabundant by a further factor of $>3$ with respect to the other elements.}
This underabundance is  probably caused by depletion into dust.
Pettini et al. (2002) took
into account the effect of dust depletion on Fe-peak elements and suggested that is of the order of a factor of two.
Matteucci \& Pipino (2002) modeled the chemical abundances of such a galaxy by taking into account the dust depletion, as suggested by Pettini et al. (2002) and concluded that this galaxy is a small young elliptical undergoing a burst of star formation and a galactic wind. We suggested an age of 35 Myr for this object. However, that original
formulation of the chemical evolution model did not take into account the dust evolution as we have here.
Further confirmation of the presence of the dust came from 3D $Ly{\alpha}$ transfer models \citep{3dly_lbg} and dust emission models \citep{dust_em_lbg}.

More recently, other lensed LBGs have been observed (e.g., Quider et al., 2009, 2010, Dessauges-Zavadsky et al., 2010) but the presence of either line emission or intervening systems, as well as the different spectral coverage, hampered the measurement of an extended set of abudance ratios as in MS1512-cB58. 
%{ Only the so-called \emph{8 o'clock arc} LBG 
%has wealth of measured abundance ratios (Dessauges-Zavadsky et al., 2010) comparable to MS1512-cB58. While the abudance pattern
%reported by Dessauges-Zavadsky et al. (2010) is very close to the one observed in MS1512-cB5, its stellar mass seems
%to be a factor of 10 larger. Therefore, we do not attempt to reproduce these two galaxies with the same model and
%postpone the analysis of the \emph{8 o'clock arc} LBG to a future work.}

For the oxygen abundance, we will also use data from the AMAZE (Assessing the Mass-Abundance redshift(Z) Evolution, Maiolino et al., 2008) and LSD 
(Lyman-break galaxies Stellar populations and Dynamics, Mannucci et al., 2009) programs at z$\sim3$ as well as data from Hainline et al. (2009).

\subsection{Quasars}
The chemical abundances in QSOs are measured either from the broad line region (BLR) or from the narrow line region (NLR). The BLR traces a region of very small size (of the order of parsecs) around the QSO, whereas the NLR extends on sizes (of the order of kpc) comparable to the size of the host galaxy.  The most remarkable facts about  abundances in QSOs are:
i) most of QSOs show oversolar metallicities, even at very high redshift($z>4$) \citep{maiolino2006}
ii) the lack of evolution in the QSOs metallicities in a redshift range $2<z<4.5$ for both BLR and NLR
(e.g. Nagao et al., 2006a, 2006b, Matsuoka et al., 2009, Juarez et al., 2009). 
The most reasonable conclusion from this is that the host galaxies of QSO formed at very high redshift and evolved very fast due to a very intense star formation which produced a lot of  metals by means of SNe II. Therefore, over-solar metallicities are attained on timescales of hundreds of Myr and after that the metallicity cannot increase anymore since the host galaxies lose their residual gas by means of galactic winds (Matteucci \& Padovani, 1993). { However, Juarez \& al. (2009) suggested that the lack of evolution can be due to selection effects.}

Recent data on NLR can be found in \cite{dodorico04}. They  derived the abundances of C, N and $\alpha-$elements from the NLR of six QSOs. We will also use these data for comparison with our models.

A non-negligible amount of dust exists in QSO and the different extinction curve between QSOs and local galaxies are challenges for models. 
%Here, we are interested in reproducing the chemical and dust mass properties of QSOs by means of a model for a very massive spheroid evolving very fast.

Here we concentrate on the quasar \qj, which is at $z \simeq  6.4$ \citep{bertoldi03b,iwamuro04}, and is one of the most distant QSOs 
detected (Fan et al. 2003). This object is only 375-800~Myr old.
Using the FIR luminosity $F_{FIR} \sim 10^{13} L_{\odot} $, the estimated dust mass
$M_{\rm dust}$ ranges from $2\cdot10^8M_{\sun}$ to $7\cdot10^8M_{\sun}$ 
(e.g. Bertoldi et al., 2003a, Robson et al. 2004; Carilli et al. 2004; Beelen et al. 2006). 

%With a
%$\sim5\times10^{10}M_{\sun}$ dynamical mass
%\citep{walter04} and an estimated $3\times10^{10}M_{\sun}$ total gas mass \citep{dwek07}
% ,one can get the a gas mass fraction  $\mu_g = 0.60$, and a dust-to-gas mass ratio  $D = 0.0067$ at $t_{g} = 375$~Myr
%%DISCUSSED THAT THE QSO ABUNDANCES CAN BE FIT AT T$>T_{GW}\sim$0.5 GYR}. 
%Since BLR is the nuclear region, the metallicity in BLR of \qj (\cite{juarez08}) 
%may be the upper limit to the the metallicity of the host galaxy.
Several studies (Barth et al. 2003; Maiolino et al. 2005; Becker et al. 2006) indicate  a near solar metallicity in the quasar host. SDSS J1148 shows a bolometric luminosity of $L_{bol}=10^{14}L_{\odot}$ powered by accretion onto a supermassive black hole (SMBH) of mass 1-5 $\cdot 10^{9} M_{\odot}$
(Willott et al., 2003, Barth et al. 2003). The estimated SFR of the host galaxy is $\sim 3 \cdot 10^{3} M_{\odot} yr^{-1}$ 
(Bertoldi et al. 2003b; Carilli et al. 2004). 
Several other high redshift quasars are now detected at submillimeter wavelengths (Priddey et al. 2003, Wang et al., 2008 and
references therein), which indicate the presence of dust grains in excess of $10^8 M_{\odot}$.

\section{Overall results}

The main parameters of the model are shown in Table~\ref{t1}.
{ In particular, the model mass is uniquely associated to effective radius $R_{eff}$, star formation efficiency $\nu$,
infall timescale $\tau$. These quantities set the predicted time for the occurrence of a wind  and the
average [Mg/Fe] in the stars. These last two values are for the central ($0-0.1 R_{eff}$), because observations typically constrain [Mg/Fe] in 
such a region. The adopted empirical relation between $\nu$, $\tau$ and the galaxy mass is motivated
by our previous works (e.g. Pipino \& Matteucci, 2004).
For each model mass, we will then consider cases where the C and O yields are taken from Maeder (1992), the N yields include primary
production in massive stars, as well as different dust recipes. These prescriptions, however, do not affect
the galactic wind timescale and the SFR.}
 Here we briefly mention some general properties of the models, such as the star formation histories (which  comply with
downsizing, e.g. Thomas et al., 2005), the dust masses and the abundance ratio evolution.
%\subsection{Cosmology effects}
%\begin{figure}
%\includegraphics[width=9cm]{lbg_age_formedz.eps}
%\caption{\label{lbg_age} }   Age at observed redshift of \lbg  \,(z=2.6) as function of the formation redshift in different cosmology model.
%\end{figure}
%\begin{figure}
%\includegraphics[width=9cm]{qso_age_formedz.eps}
%\caption{\label{q_age_z} }  Age at observed redshift of \qj  \,(z=6.4) as function of the formation redshift in different cosmology model.
%\end{figure}
In Pipino \& Matteucci (2004) the mean formation redshift of the model galaxies was constrained to the observed ages (inferred from optical
line-strength indices) and the mean colours at a given mass. For the purpose of this paper, we leave the 
formation redshift free in order to find the best fit to the single objects that we try to reproduce.
%It is worth noting that 
%the best agreement we find for \lbg \, is obtained by assuming a formation redshift z=3.2 in the standard LCDM cosmology, 
%whereas for  the QSO J1148+525 the best agreement is obtained with a formation redshift $z>8$. 
%Indeed,
%LBGs seem to be small ellipticals and that they formed after the big spheroids hosts of QSOs for which we need a formation redshift z=8
%In Figure 3 we show the predicted evolution of the gas mass for a galaxy of $M=10^{12}$, which can 
%be directly compared with QSO SDSSJ1148.

\subsection{Star formation rates}

{ Fig.~\ref{sfr} shows the predicted SFR for our model galaxies summed over the shells and
truncated at the average $t_{gw}$. It is worth noting that the average 
value of $t_{gw}$ is lower than that calculated for the inner regions and 
reported in table 1. 
The reason is that, in the adopted framework, elliptical
galaxies form outside-in (e.g. Martinelli et al., 1998, Pipino et al., 2008), hence
a wind developes first in the outskirts, whereas the innermost regions
are still experiencing star formation (and QSO growth).}
The star formation efficiencies
have been chosen to reproduce the observations - including those in local ellipticals - and are 
not considered as free parameters in the remainder of the paper.
%Here we show the results for two fiducial galactic baryonic masses: $3 \cdot 10^{10}$ and $10^{12}M_{\odot}$.
As discussed also in Matteucci \& Pipino (2002) and it is also clear from Fig.~\ref{sfr}, as
far as the SF histories are concerned, a model with mass in the range $10^{10}-10^{11} M_{\odot}$
is the only one which must be compared to MS1512-cB58, whereas the most massive galaxy should be compared to the host of the QSO \qj.
In particular,  the star formation rate evolution in Fig.~\ref{sfr} shows the \emph{downsizing}, 
namely that massive galaxies have more efficient star formation with a shorter duration  than less massive galaxies. 
%In the Figure it is also indicated the SFR derived for the LBG galaxy \lbg \, and the
%QSO host \qj. %Best agreeement is for the models with mass $3 \cdot 10^{10}$ and $10^{12}M_{\odot}$. 

\begin{figure}
\label{sfr}
\includegraphics[width=9cm]{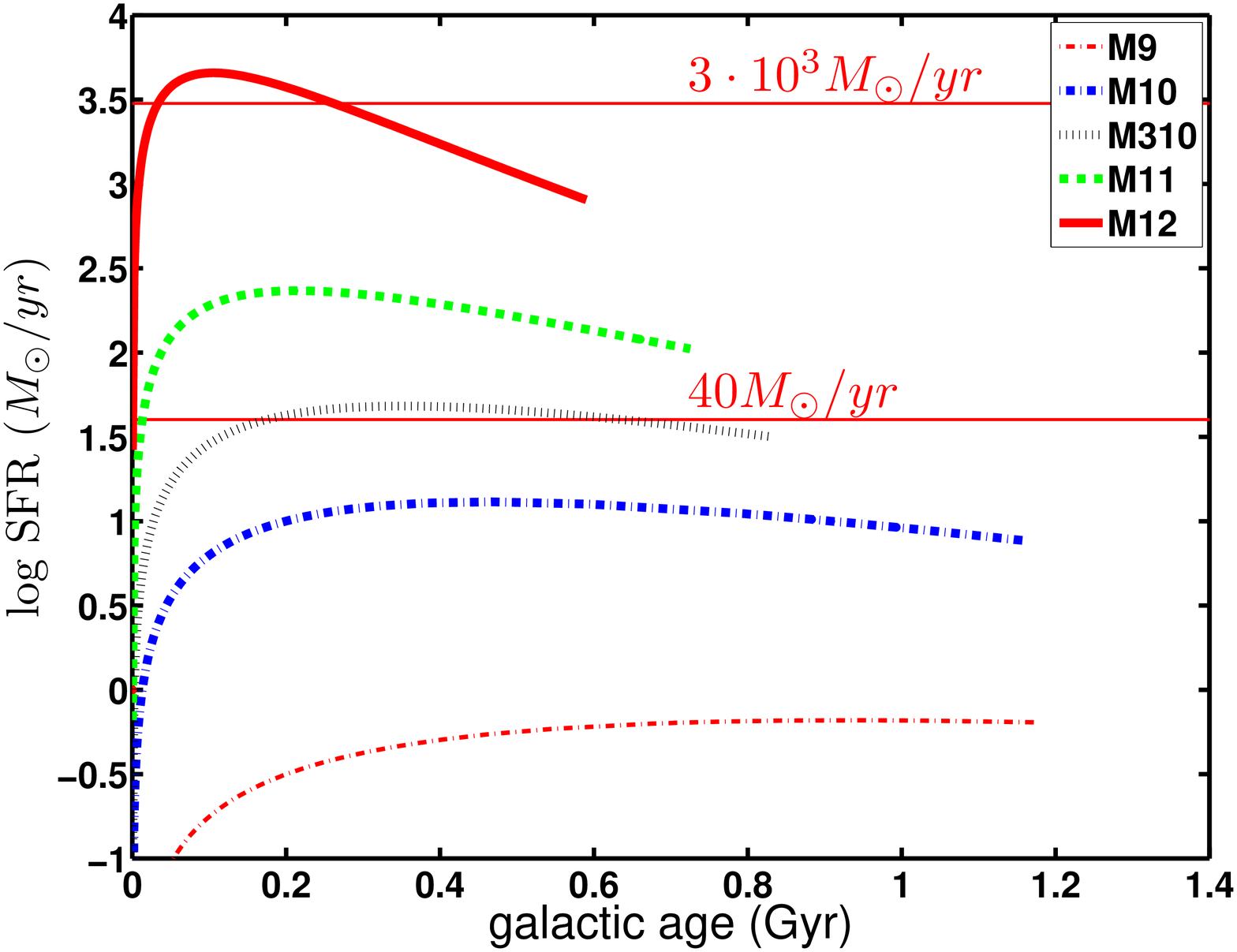}
\caption{\label{sfr} The predicted star formation rate as function of  galactic age for the model galaxies. Note that the setup
for the dust and adopted yields do not modify the predicted star formation rate. Therefore each line in the plot is representative
of all models with a given mass. The horizontal lines correspond to a SFR=$40 M_{\odot}yr^{-1}$, which is the estimated value for \lbg \citep{pettini02_cb58},
and to SFR=$3\times10^3 M_{\odot}yr^{-1}$, which is the value inferred for \qj , respectively.
}
\end{figure}

\subsection{Dust masses}

\begin{figure}
\includegraphics[width=9cm]{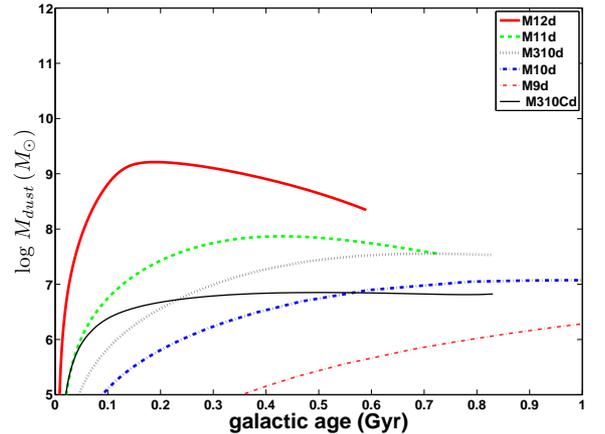}
\caption{\label{dust6} The evolution of the dust mass in the galactic models.}
\end{figure}

\begin{figure}
\includegraphics[width=9cm]{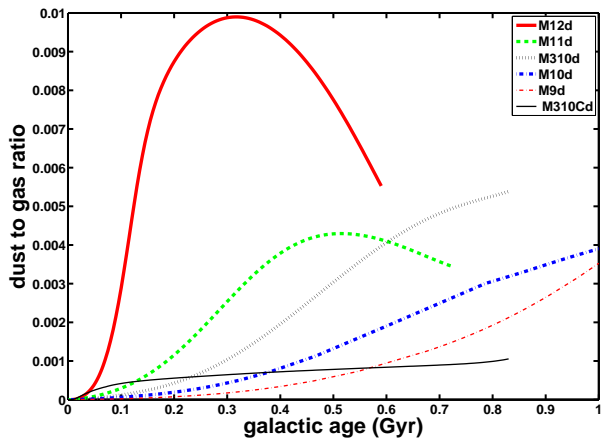}
\caption{\label{dustg6} The evolution of the dust to gas ratio in the galactic models. }
\end{figure}

{ The evolution of the dust mass and of the dust-to-gas ratio are shown in Figs.~\ref{dust6} and~\ref{dustg6},
respectively. A dust mass-galaxy mass relation, namely that dust content is higher and
increases faster in more massive galaxies, is evident. 
{ A linear regression analysis yields the following relations 
\begin{equation}
Log (M_{dust,peak}/M_{\odot})=0.98 (Log M_{lum}/M_{\odot}) - 2.75 \,
\end{equation}
when the dust mass at the peak is considered, whereas
\begin{equation}
Log (M_{dust,05}/M_{\odot})=1.07 (Log M_{lum}/M_{\odot}) - 4.01 \,
\end{equation}
after 0.5Gyr of evolution.}
Such a relation stems from the SFR-mass relation
in Fig.~\ref{sfr} and it is likely to present a redshift evolution similar to the observed
one in the mass-metallicity relation (Maiolino et al., 2008).
This is a clear and novel prediction of our set of models. Future observations can provide
a confirmation of this picture. 
For the $3 \cdot 10^{10} M_{\odot}$ mass case we also show the dust mass which 
would be obtained by using the prescriptions by Calura et al. (2008, thin solid line) 
compared to a model with same mass, but the new prescriptions for the dust (dotted line).
This latter model will be considered the fiducial case to study the galaxy \lbg.
The Calura et al. prescriptions yield a faster and earlier increase in the dust mass, due to the larger
$\delta_i^{II}$. At later times, however, the effect of the stronger destruction term hampers 
further increase in the dust mass as opposed to the model with the new prescriptions.
The implications of such a different behavior will be further discussed below by means of the chemical
abundance ratios.}

\subsection{Abundance ratios in the gas}

\begin{figure}
\includegraphics[width=9cm]{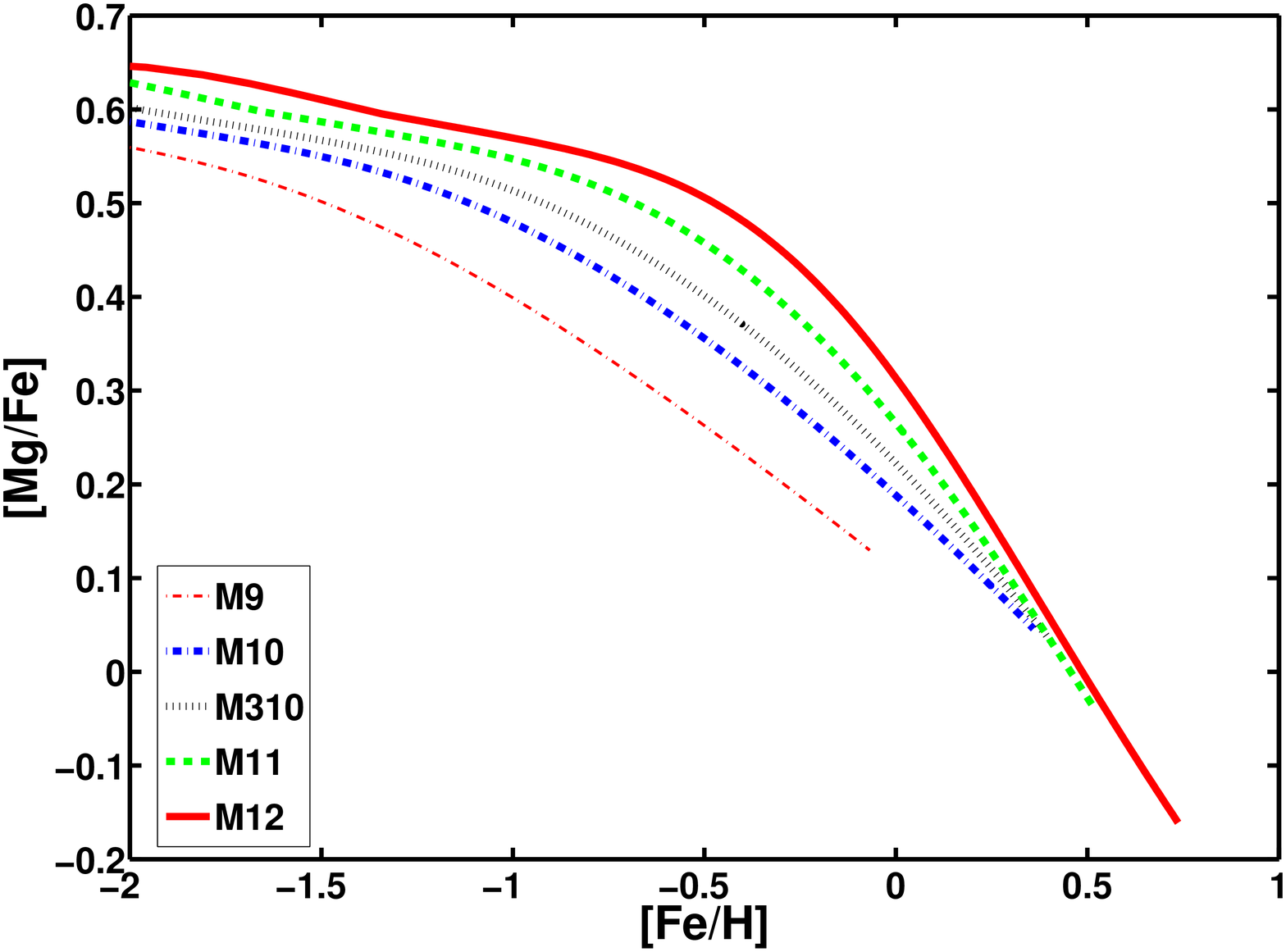}
\includegraphics[width=9cm]{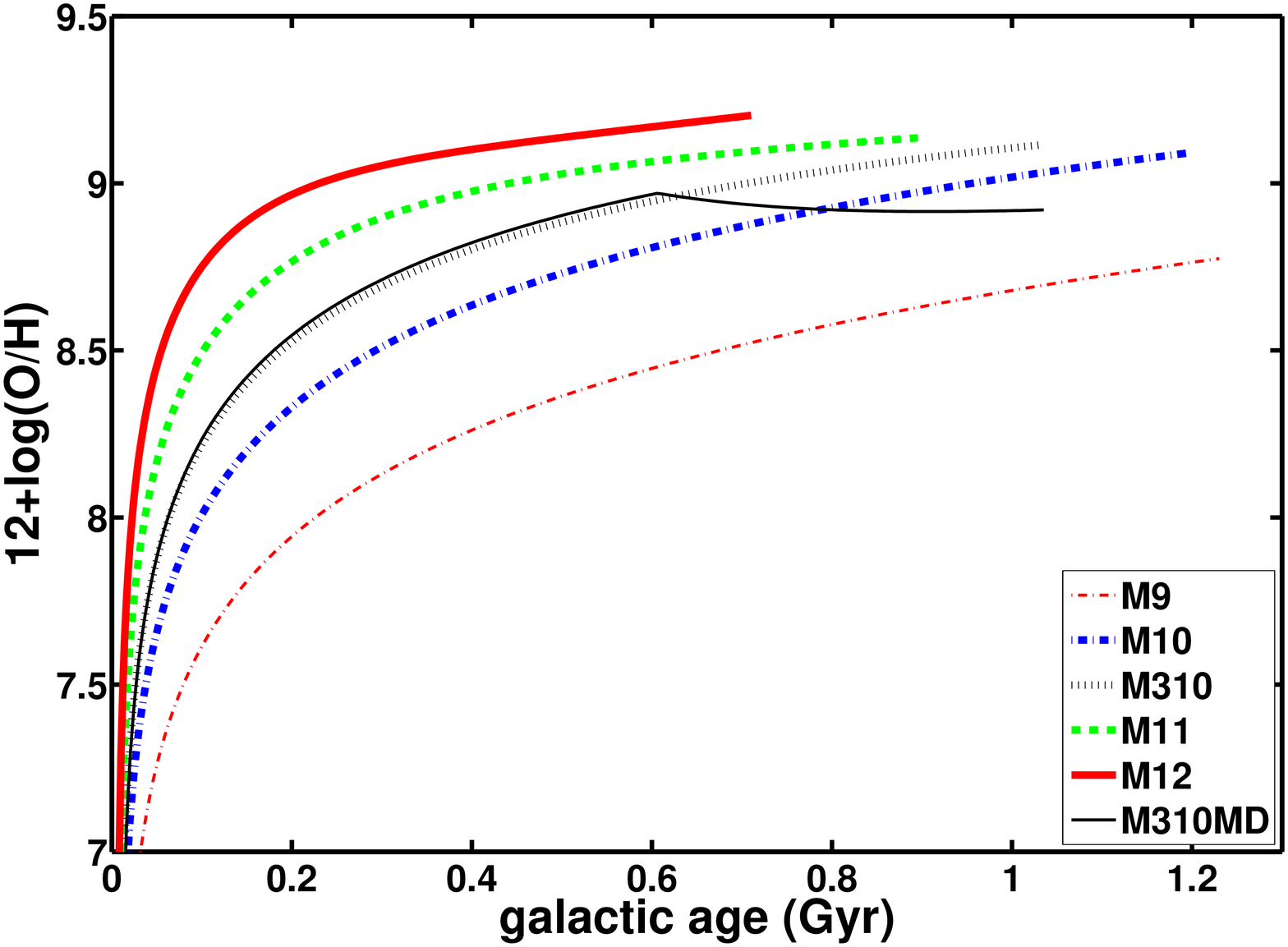}
\caption{\label{allMgFe} The predicted [Mg/Fe]-[Fe/H] relation for models with different masses (upper panel)
and the evolution of the metallicity in terms of the O abundance (lower panel).}
\end{figure}

{  
The different timescales of the enrichment processes for  
$\alpha-elements$ (massive star ending as SNe II),  
and iron (dominated by intermediate mass stars in binary systems ending as SNe Ia) imply an enrichment delay of iron
and this explains the evolution of all curves in Fig.~\ref{allMgFe} (upper panel).
If we then compare the various models in Fig.~\ref{allMgFe} (upper panel) we can see the effect of downsizing very well. 
Due to the relation between star formation efficiency $\nu$, infall time scale $\tau$  and the total luminous  mass $M_{lum}$ mentioned in \S 2 , 
we predict that the [Mg/Fe] ratio in the gas out of which stars form is higher for more massive galaxies.
This ensures the fit of the [$\alpha$/Fe] versus mass for ellipticals, as shown by Pipino \& Matteucci (2004).
Moreover (Fig.~\ref{allMgFe}, lower panel), the increase in the [X/H] with time is much faster and saturates
at higher (over-solar) values in the more massive galaxies than in low mass ones.
This ensures that, despite more massive galaxies have less time to convert gas into stars, they do it more
efficiently than lower mass systems, thus reproducing the mass-metallicity relation that we observe
in the stars of local ellipticals, namely a increase of a factor of $\sim$3 in the metallicity
in the mass range $10^{9}-10^{12} M_{\odot}$.
Until recently, only the mass-stellar metallicity relation (e.g. Thomas et al., 2005) could be measured,
namely a relation that is inferred from the integrated light of local ellipticals. While important
in constraining the galaxy formation process, such a measure reflects an average metallicity 
and, hence, does not allow to capture the details of the star formation histories as opposed
to the study of the, e.g., O/H ratio versus time.
Very recent observations (e.g. Maiolino et al., 2008), provided the opportunity to directly compare
model predictions with gas abundances of z$\sim$3 star forming galaxies.
Indeed Calura et al., (2009) find that the observed mass-gas metallicity relation at such high
redshift can be explained only by models of elliptical galaxy progenitors as the ones studied here.
Here we do not repeat the analysis. Moreover, we show all the models as if they were forming at the same time.
In principle, both panels of Fig.~\ref{allMgFe} can be used as a diagnostic plot for high redshift
galaxies: the measurement
of the chemical abundance of one element which is not easily dust depleted (like O) and
the abundance ratio of two elements which have very similar condensation efficiencies (like Mg and Fe)
can immediately tell the age and the mass of the galaxy as well as the past SF history, hence
the likely morphology at z=0.
%There are evidences (e.g. Thomas
%et al., 2005) that more massive galaxies are on average older than low mass ones, and we will see
%below that this also applies also comparing the single objects, namely \qj \, to \lbg.
%If one takes this into account, the evolution in the mass metallicity relation is faster and
%more ... in the end regime 
}

\begin{figure}
\includegraphics[width=9cm]{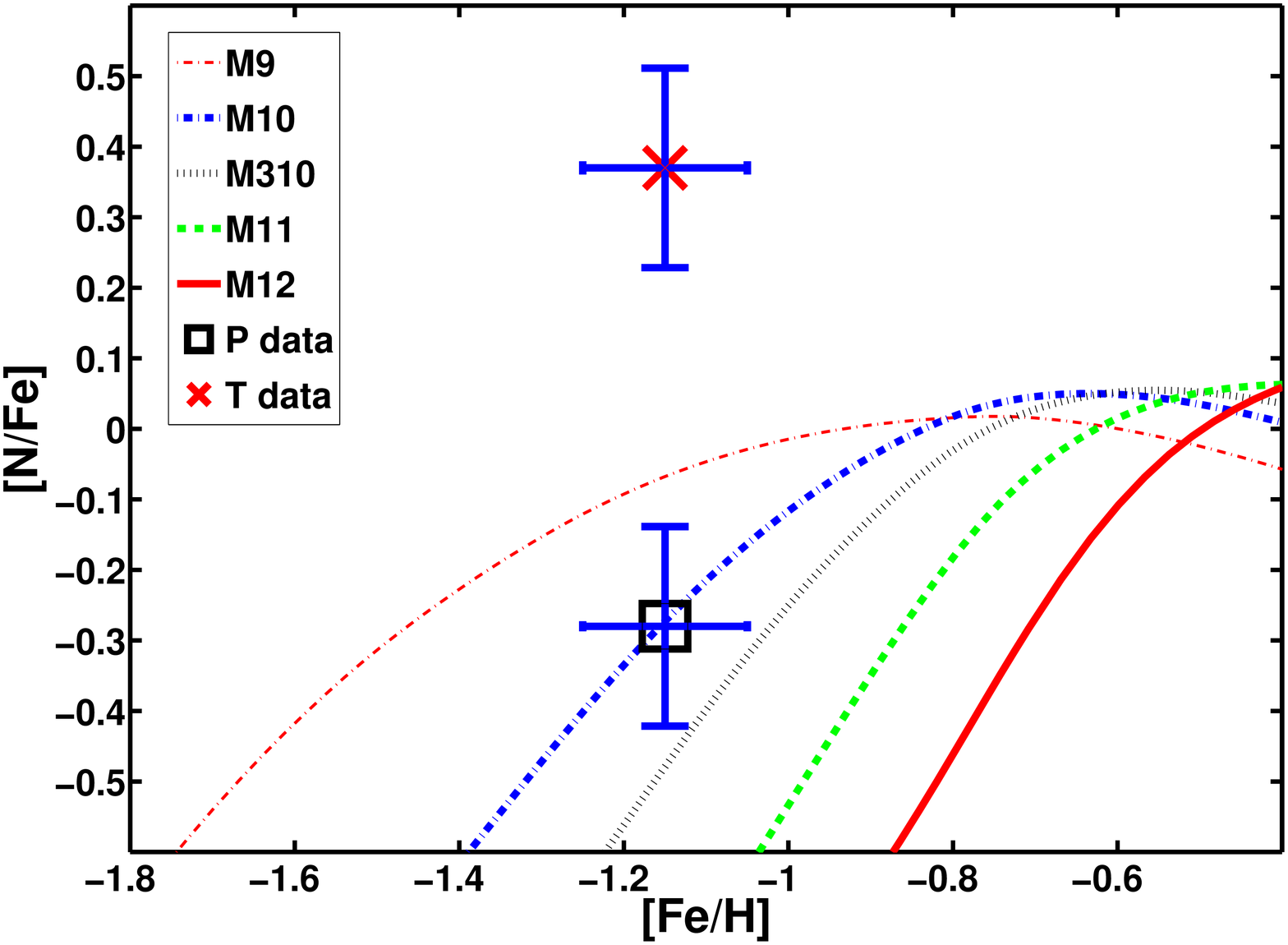}
\includegraphics[width=9cm]{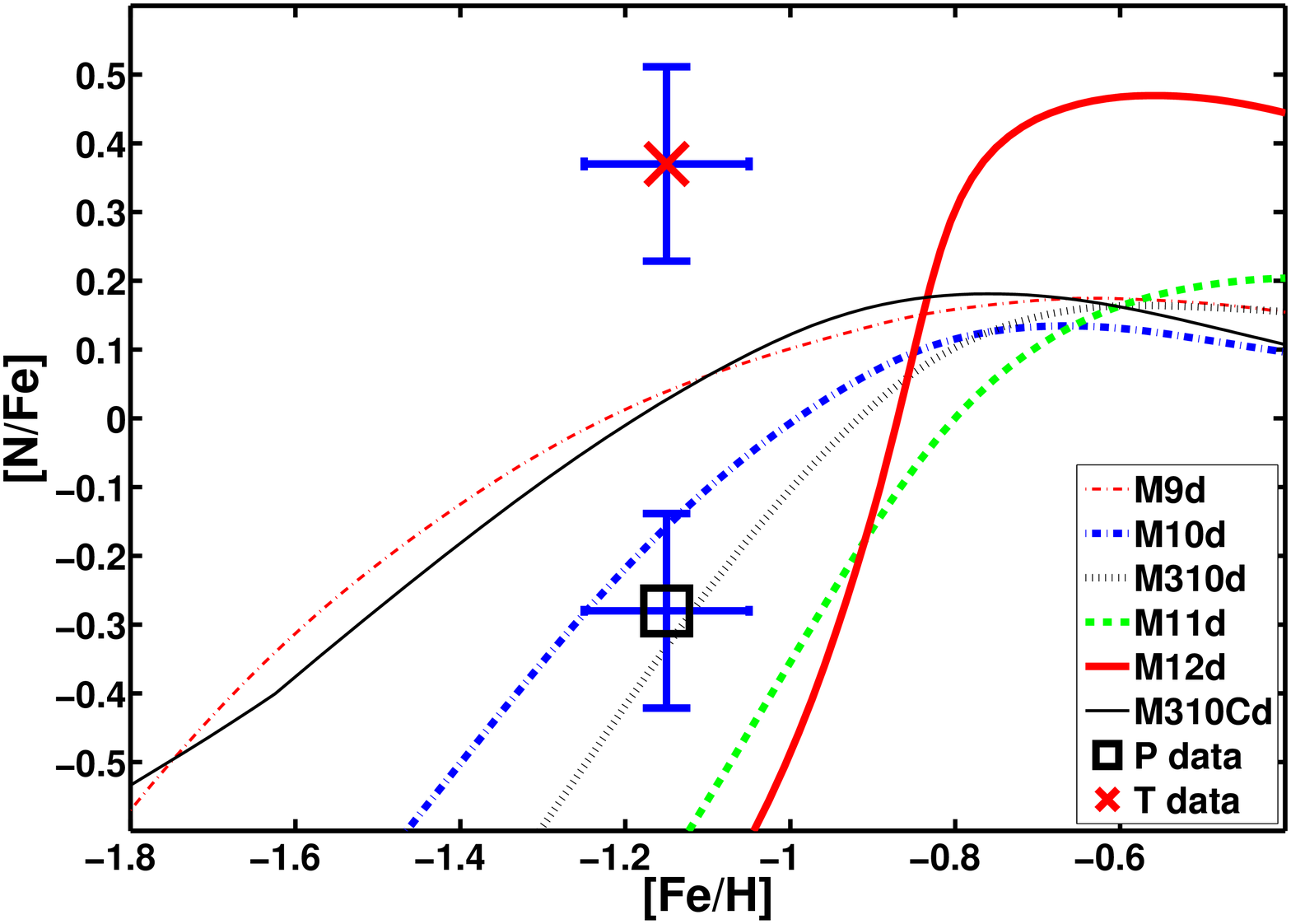}
\caption{\label{allNFe} The predicted [N/Fe]-[Fe/H] relation for models with different masses (with and without dust:
lower and upper panel, respectively).
Data for \lbg \,are taken from \cite{teplitz00_cb58} (Red Cross, T data), \cite{pettini02_cb58}(Black Square,P data).}
\end{figure}

\begin{figure}
\includegraphics[width=9cm]{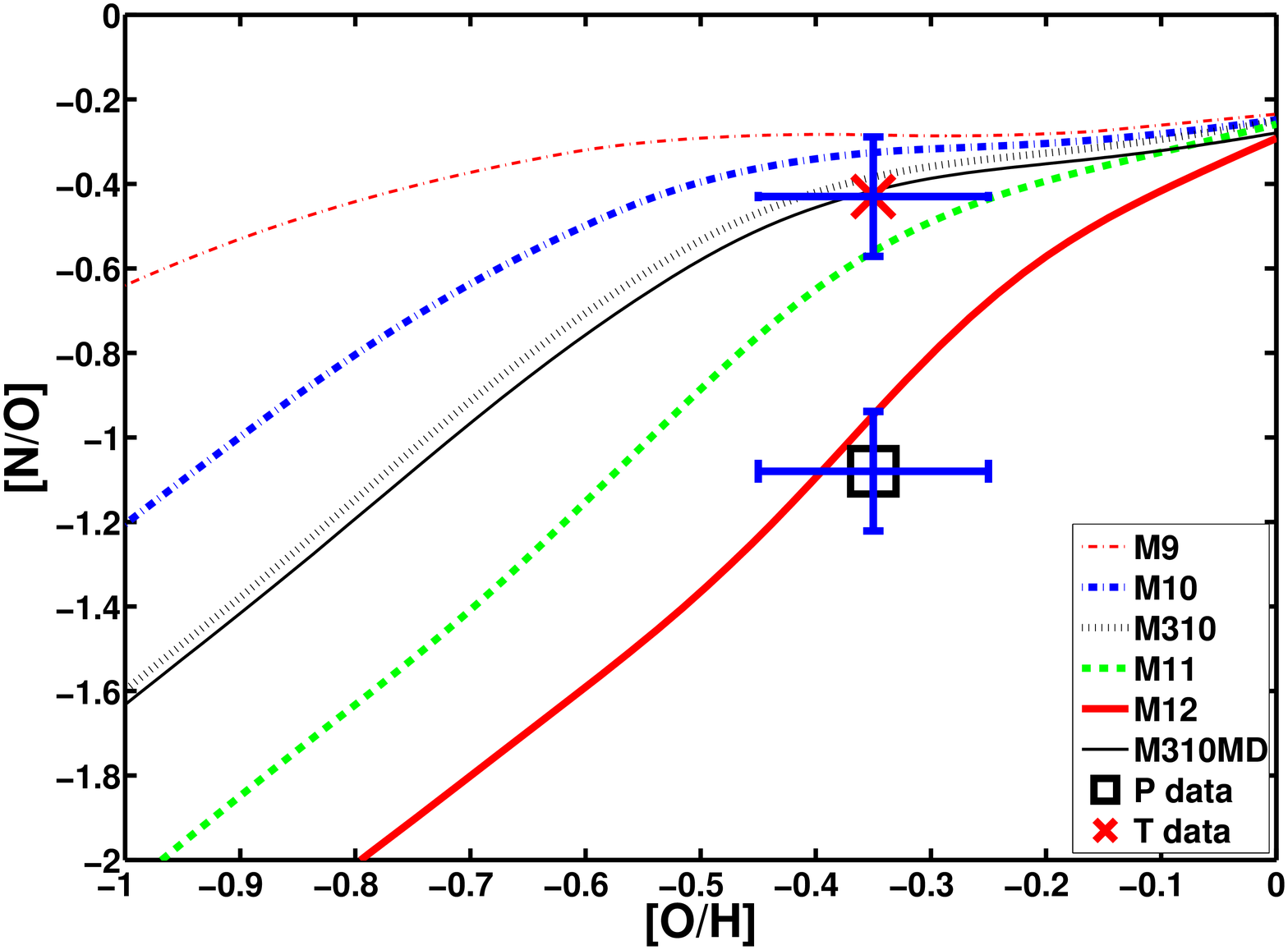}
\includegraphics[width=9cm]{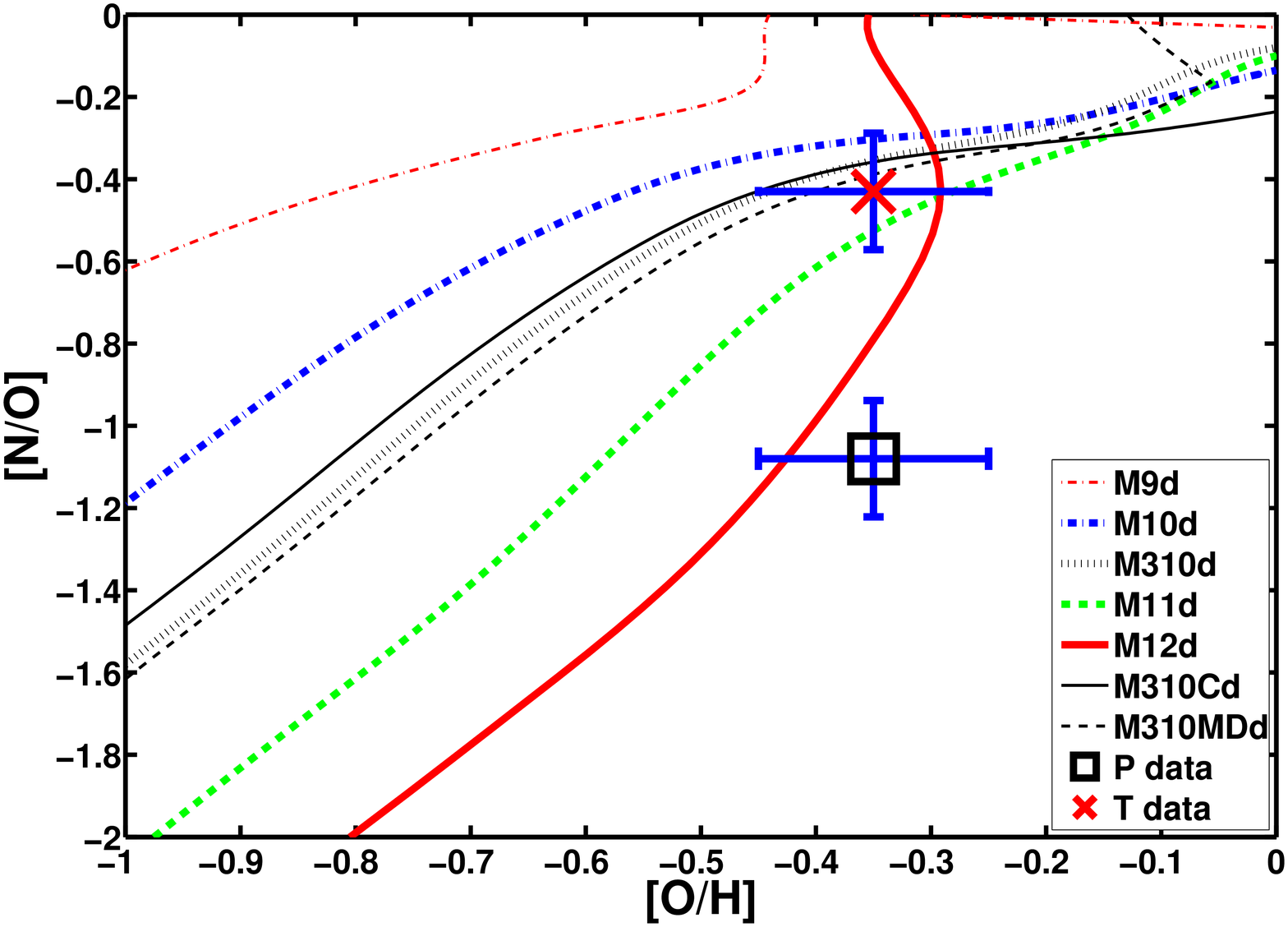}
\caption{\label{allNO} The predicted [N/O]-[O/H] relation for models with different masses (with and without dust:
lower and upper panel, respectively). 
Data for \lbg \, are taken from \cite{teplitz00_cb58} (Red Cross, T data), \cite{pettini02_cb58}(Black Square,P data).}
\end{figure}

{ Finally, we show the [N/Fe] (Fig.~\ref{allNFe}) and [N/O] (Fig.~\ref{allNO}) ratios as a function of the metallicity
for galaxies of differing masses. The predictions are made for models with (lower panels) and without dust (upper panels). 
Here we anticipate from the next section the observed values in the galaxy \lbg \, , since
they represent a valuable reference point to guide the eye. These data 
are taken from \cite{teplitz00_cb58} (Red Cross, T data, emission lines), \cite{pettini02_cb58}(Black Square,P data, absorption
lines). Only the Pettini et al. data have been used as constraint in our previous works.
Here we assume that the two datasets limit the observed region and the [N/Fe]
from Pettini et al. as a lower limit which must be necessarily reproduced by the models.
From a joint analysis of Figs.~\ref{allNFe} and~\ref{allNO} we infer that only models
with mass \emph{lower} than $3 \cdot 10^{10} M_{\odot}$ predict the [N/Fe] within the observed range
at [Fe/H]=-1.15. On the other hand, only models with mass \emph{larger} than $3 \cdot 10^{10} M_{\odot}$
predict values for the [N/O] ratio consistent with observations.
Therefore, in agreement with what has been inferred from the SFR analysis (Fig.2), 
we will focus on a $3 \cdot 10^{10} M_{\odot}$ model to study the galaxy \lbg .
For this model, the predictions based on the Calura et al. (2008) prescriptions are also shown.
Since N is not depleted into dust and O has a very small depletion, both the new
and the Calura et al. prescriptions lead to basically identical results in Fig.~\ref{allNO}.
On the other hand, the change is more evident in Fig.~\ref{allNFe}. As seen above, in
fact, the Calura et al. prescriptions give more dust at earlier times with the
net effect of increasing the predicted [N/Fe] ratio by 0.3 dex.
At [Fe/H]$\sim -$1.15, nearly 50\% of the Fe is locked into dust following the Calura et al. prescriptions,
whereas the fraction is only 12\% with the new ones.

Here it is worth mentioning that
[Fe/H] = -1.15 is attained at 0.05 Gyr in model M11, and at 0.15 Gyr in the M10 case.
In other words, at a fixed metallicity [Fe/H] = -1.15, the model M11 has [N/H]$ < -2$,
whereas the model M10 has [N/H] $\sim -0.8$, hence a much higher [N/Fe]. This happens because the smaller galaxy takes a longer time to 
attain the same metallicity. A similar explanation holds for the [N/O]-[O/H] diagram. 
It can be shown that this is a consequence of assuming $\nu$ increasing with mass (i.e.the downsizing), 
whereas, if one assumes that $\nu$ decreases with mass as in the original monolithic collapse model
by Matteucci \& Pipino (2002), more massive galaxies will exhibit systematically higher [N/O] values than low mass ones
at a fixed [O/H] (see, e.g., Figs. 2 and 4 in Matteucci \& Pipino, 2002).
Such a difference in the mass dependence of the predicted abundance ratios trends with metallicity
is a non trivial effect of downsizing that, to our knowledge, has not been shown elsewhere
and still awaits confirmation by observations.

We remind the reader should that the uncertainties in the stellar yields (e.g. Romano et al., 2010) affect the galaxy properties,
e.g. the age, that we infer when comparing models to observations and indeed it is difficult
to assign a quantitative measure (i.e. an error) to this effect.
For instance, since low mass models have a subsolar gas metallicity 
for most of their evolution, we expect a negligible impact of the mass-loss
dependent O yield which seem to be required to explain the abundance pattern in metal rich systems as the Galactic bulge
(McWilliam \& al. 2008; Cescutti \& al. 2009).
Indeed, as it can be seen from Figs.~\ref{allMgFe} and~\ref{allNFe}, a $3 \cdot 10^{10} M_{\odot}$ 
mass model which differs from the fiducial case only for the C and O yields, modified
according to Maeder (1992), does not predict any significant departure in the abundance
ratios { in the metallicity regime which is relevant for our study. We do see, however,
a $\sim$0.2 dex lower O abundance (lower panel of Fig.~\ref{allMgFe}) with respect to the fiducial
case at times later than $\sim$0.6 Gyr. This happens when the metallicity exceeds the solar value ($\sim$8.7 - Asplund
et al., 2005 - in the units
used in the figure) and the strong mass loss in the Maeder (1992) yields does not allow the O abundance
to increase futher.}
We will see that this is not the case for the most massive and metal rich objects in Sec. 7.
In the next section we show some examples of modifications to N yields.
Here we note that not only the yields, but also the prescriptions for binary stars 
which explode as SNIa may affect our results.
We checked that, given the relatively short time needed to attain [Fe/H]$\sim$-1.2 in the models, most of the Fe
has been produced by SNII and changes in the SNIa
prescriptions do not really affect the predicted abundance ratios.
This can be easily verified by looking at the upper panel of Fig.~\ref{allMgFe}: all the models
are still in the \emph{plateau} phase. 
Therefore the results in this age/metallicity regime that we will present in Sec. 6 are robust against changes in the SNIa prescriptions.
However, we expect more relevant effects at later ages/larger [Fe/H].
Adopting recent observationally motivated prescriptions for both
\emph{prompt} and \emph{late} SNIa as in Mannucci et al. (2006),
the changes in the abundance ratio evolution as a function of [Fe/H] are minimal
(c.f. Fig. 8 in Matteucci et al., 2006).
If instead, one assumes the Double-Degenerate scenario (Iben \& Tutukov, 1984) as opposed
to the Single-Degenerate one adopted here, recent work by Greggio (2005) and Valiante et al. (2009b) show that
the expected number of SNIa event in the former case, and hence the Fe produced, is much lower than
in the latter. Note that, in order to make a meaningful comparison, 
all the different channels must be constrained to reproduce the observed present-day SNIa rate. }

\section{The low mass case: comparison with LBGs}

In this section we focus on our low mass model in comparison to the class of the LBG galaxies.
As noted in the Introduction, MS1512-cB58 is still the best candidate for having an extesive sample of well measured abundance ratios. Therefore, this galaxy will is the best case to test our model.

Absolute abundance evolution usually depends on all the chemical evolution model assumptions, whereas the abundance ratio evolution only depends on stellar yields, stellar lifetimes, and IMF. Therefore, abundance ratios can be used as cosmic clocks if they involve two elements enriching the ISM on quite different timescales, such as  $[\alpha / Fe]$  and $[N/ \alpha]$ ratios. { These ratios, when plotted as functions of metallicity tracers such as [Fe/H], 
allow us to clarify the particular star formation history (see Matteucci 2001)}. 
%Since some elements (e.g. Si, Fe) are depleted more into dust than others (e.g. O, Zn), 
%it is interesting to study the evolution of  those different  elements.

\subsection{[N/O] and [N/Fe] ratio}

\begin{figure}
\includegraphics[width=9cm]{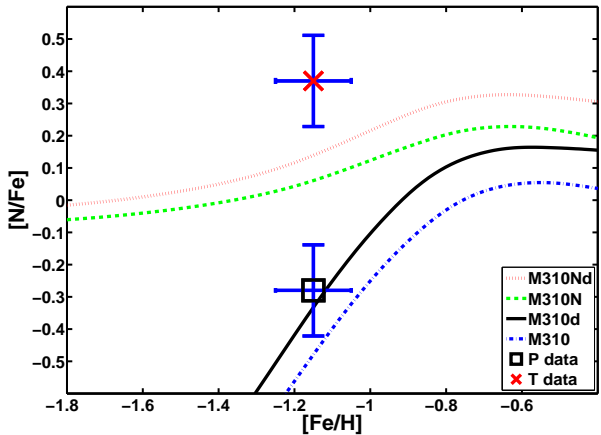}
\caption{\label{nfinfall} The predictions of [N/Fe] vs [Fe/H] ratios for a $3 \cdot 10^{10}$ model galaxy.
Cases with either primary+secondary or secondary only N production in massive stars are shown along
with models with and without dust depletion. 
Data are taken from \cite{teplitz00_cb58} (Red Cross, T data), \cite{pettini02_cb58}(Black Square,P data).}
\end{figure}

%\begin{figure}
%
%\includegraphics[width=8cm]{xray_N_Fe_noinfall.eps}
%\caption{\label{nfnoinfall} The predictions of [N/Fe] vs [Fe/H] ratios for different mass galaxies without infall case. Massive galaxies have less [N/Fe], which caused by downsizing effect. Data are taken from \cite{teplitz00_cb58} ($*$, T data),\cite{pettini02_cb58}($+$,P data)
%}
%\end{figure}

\begin{figure}
\includegraphics[width=9cm]{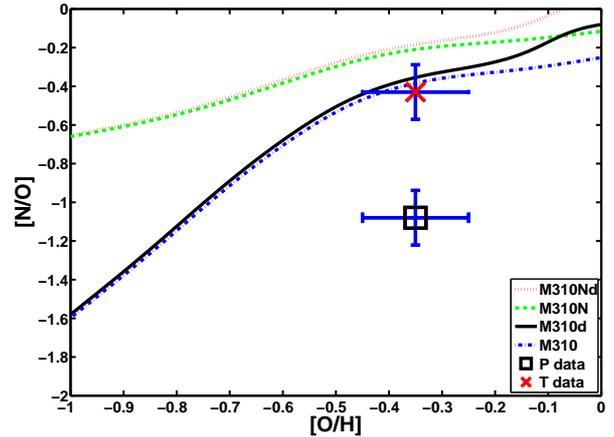}
\caption{\label{no_infall} The [N/O] vs [O/H] trend. 
Cases with either primary+secondary or secondary only N production are shown along
with models with and without dust depletion. 
Data are from \cite{teplitz00_cb58} (Red Cross, T data), \cite{pettini02_cb58}(Black Square, P data).}
\end{figure}

As shown by Matteucci \& Pipino (2002), it is useful to study the abundance ratio of a refractory 
element (like Fe) to
an undepleted one (like N) in order to measure the amount of dust depletion for the Fe-peak elements.
In Fig.~\ref{nfinfall} we show the predictions from the $3 \cdot 10^{10}M_{\odot}$ galaxy model
with different prescriptions regarding the dust and the N primary production.
%The data for \lbg \, are from Pettini et al. (2002) (N, Fe measured from absorption lines)
%and from Teplitz et al. (2000) (N measured from emission lines). We refer to Pettini et al. (2002)
%for a thorough discussion between the two observational values. Here we consider their difference
%as an estimate of the systematic errors involved in the measurements, and the [N/Fe]
%from Pettini et al. as a lower limit which must be necessarily reproduced by the model.
Fig.~\ref{nfinfall} shows that in order to predict a value for [N/Fe] in the observed range, 
dust depletion is needed, although such a conclusion suffers from a degeneracy  with the
N production. 
Nitrogen is generally considered mainly a secondary element, in the sense that N needs the C and O,
originally present in the star, to be created during the CNO burning cycle.
However, N can also be produced as a primary element in some special situations, 
such as in AGB stars (Renzini \& Voli, 1981; Van den Hoeck \& Groenewegen, 1997; Marigo 2001)  
and massive rotating stars (Meynet \& Maeder, 2002).
In fact, the C and O going to form N are not the original ones present in the gas out of which the star formed, 
but they have been synthesized  by the star itself. In the two different cases N production changes significantly, 
especially at low metallicities.
From a look at Figure 9 we can derive the following considerations: it is difficult to say 
whether the N is primary or secondary in massive stars. 
First of all because Fe is depleted into dust and second because of the discrepancy
between the N measured from emission and absorption lines. 
On the other hand, Figure 10 can give more indications. 
In particular, it seems that assuming primary N from massive stars overproduces N
with respect to O in the gas { by more than a factor of three. On the other hand, the model
which best reproduces the [N/Fe] value from Pettini et al. (2002), namely the fiducial
model with dust and without primary N, is offset by Pettini et al.'s [N/O] ratio
by 0.8 dex}.
Here we add that at [Fe/H]=-1.15 the model has an age of 100 Myr, therefore
we confirm that \lbg \, must be young. %At such an age the fraction of Fe in dust is 12\%.

\subsection{$\alpha-elements$ to iron ratios}

\begin{figure}

\includegraphics[width=9cm]{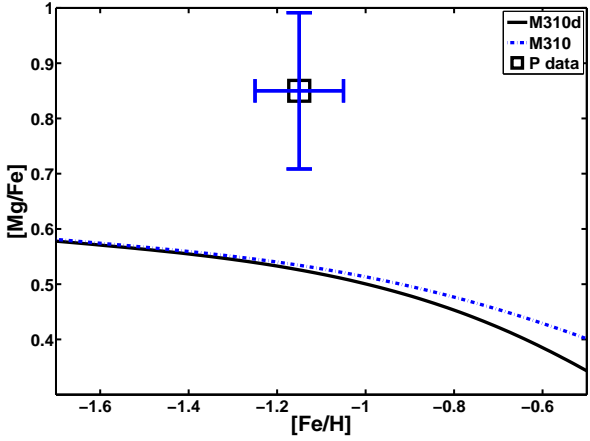}
\caption{\label{mgfe} The [Mg/Fe] vs [Fe/H] ratios  for the $3 \cdot 10^{10}M_{\odot}$ galaxy model. Data from Pettini et al. (2002).}
\end{figure}

\begin{figure}
\includegraphics[width=9cm]{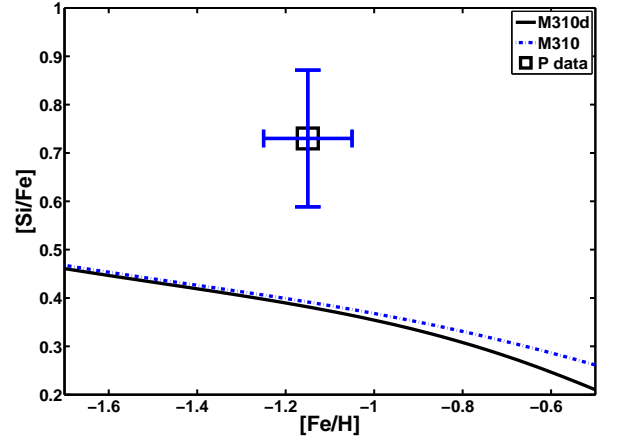}
\caption{\label{sife} The[Si/Fe] vs [Fe/H] ratios for the $3 \cdot 10^{10}M_{\odot}$ galaxy model. Data from Pettini et al. (2002).}
\end{figure}

Fig.~\ref{mgfe} and Fig.~\ref{sife} show the predicted [Mg/Fe] and [Si/Fe]  vs. [Fe/H] compared to observational 
data for the LBG \lbg \, (Pettini et al. 2002). Our predicted [Mg/Fe] and [Si/Fe] ratios are still 
lower than observed in \lbg . On the other hand,  the predicted values are higher by $\sim$ 0.2 dex than 
predicted by Matteucci \& Pipino (2002), { because the yields adopted in this work consider a sligthly higher Mg production from massive stars with respect to the Woosley \& Weaver (1995) yields (namely the yields adopted by Matteucci \& Pipino, 2002).}
Note that in the so-called ``horseshoe'' LBG (Quider et al., 2009),  the
Si/Fe ratio is very close to the one observed in cB58, thus confirming the discrepancy between theory
and observations.
We do not believe that the discrepancy could indicate a problem related to too low Mg and Si yields in the adopted nucleosynthesis, since
a [Mg/Fe] $\ge$ 0.6 dex in the gas for a large fraction of the galaxy evolution 
would imply [$<$Mg/Fe$>_{stars}$] $>$0.3-0.4 dex, which exceeds the observed value observed in present day low mass ellipticals. 

{ When the metallicity is measured from the photospheric absorption features of the stellar light (Rix et al., 2004), instead, the 
[Fe/H] is 1.5-4 times higher than those reported by Pettini et al. (2002), hence}
there is the possibility that this disagreement originates from the dust, i.e. a possibility could be that in MS 1512 -cB58 Si 
and Mg are not as depleted as much as Fe, at variance with our prescriptions. 
{ For instance, if we leave $\delta_{Mg}^{II}=\delta_{Si}^{II}$=0.08 as prescribed by the new model presented
in this paper and increase $\delta_{Fe}^{II}$ to 0.5, we predict [Mg/Fe]=0.75 and [Si/Fe]=0.6 at [Fe/H]=-1.15, hence formally in agreement
with the observed value, within the error.}
This, however, cannot be justified on theoretical 
grounds since we assume the same efficiencies for Mg, Si and Fe because they have nearly the same condensation temperature 
(see Calura et al.2008 for details). { However, such a differential depletion effect, which is known to affect the abundance pattern of the cold gas in  
 Damped Lyman Alpha systems (Vladilo 2002, Calura et al. 2003), 
could partially alleviate the discrepancies between models and observations.}
The difference between predicted and observed values is not so evident in the Si/C abundance in QSOs (see below).

\subsection{The O/H ratio and the age of LBGs}

\begin{figure}
\includegraphics[width=9cm]{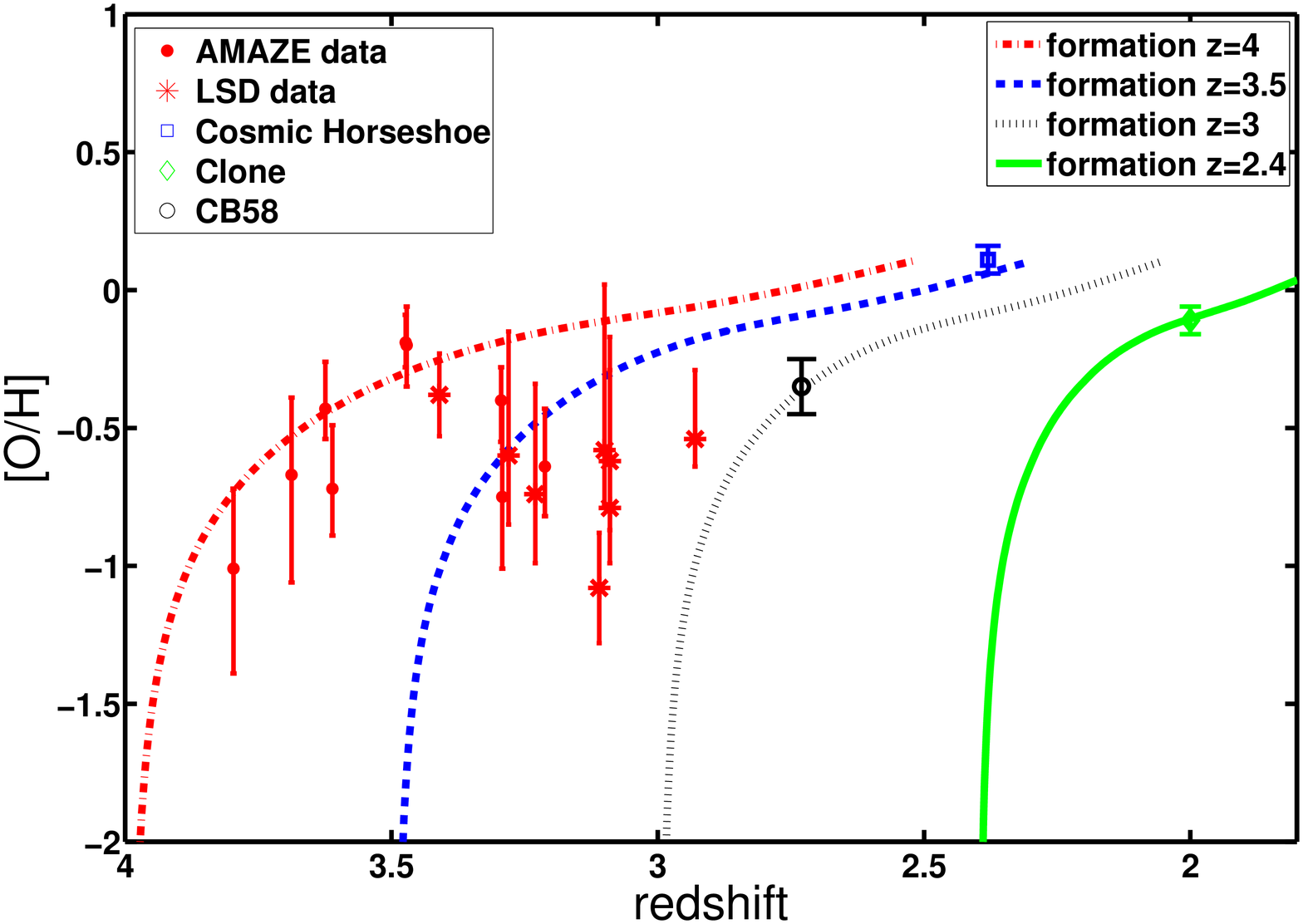}
\caption{\label{no_gtinfall} [O/H] ratio as a function of redshift for a $3 \cdot 10^{10}M_{\odot}$ galaxy and different case
formation redshifts. Data are from Maiolino et al. (2009, AMAZE), 
Mannucci et al. (2009, LSD), \cite{pettini02_cb58}(\lbg). Clone and Cosmic Horseshoe are from Hainline et al. (2009).}

\end{figure}

{ Finally, in Fig.~\ref{no_gtinfall} we show the predicted age-oxygen relation for Lyman-break galaxies, compared with the recent data from Maiolino et al. (2009, AMAZE), Mannucci et al. (2009, LSD), Hainline et al. (2009) and Teplitz et al. (2000).} Here we obtain a good fit to the data if we assume a galaxy formation redshift for these galaxies in the redshift interval $\sim$2.3-4.
%in agreement with what concluded before for  the galaxy \lbg \, which is also reported in the figure.
Note that all the galaxies have masses around a few 10$^{10}M_{\odot}$, therefore the inferred spread in the formation
epoch can be understood as the redshift range in which low mass ellipticals form.
{ In particular, the curve which gives the best agreement with \lbg \, data in Fig.~\ref{no_gtinfall} requires a formation redshift z$\sim$3, that
is, in the standard LCDM cosmology, an age of at least 0.2 Gyr (we recall that \lbg is at z=2.73). A similar conclusion can be reached from the analysis
of Fig. 2, where the best agreement with the observed SFR is obtained at ages larger than 0.2 Gyr. However,
given the uncertainties in the observational values and the flat trend with time in the predicted one,
such a constraint is rather weak. Finally, we showed above that the time at which our model reaches
the metallicity [Fe/H]=-1.15 is 0.1 Gyr, again hinting for a very young age for the galaxy \lbg \, . }

\subsection{Abundance ratios in the dust}

\begin{figure}
\includegraphics[width=9cm]{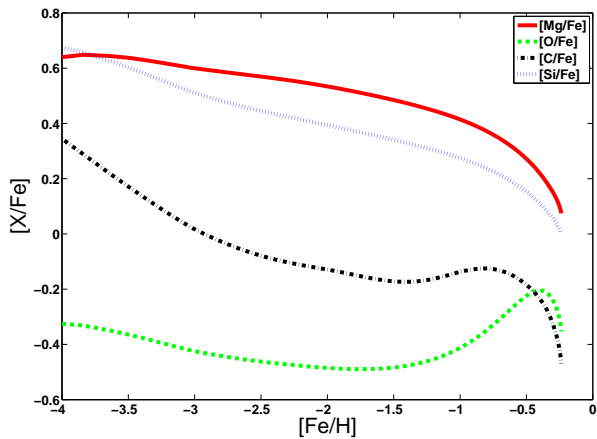}
%\includegraphics[width=9cm]{x_fe_Fe_h_in_dust_M12d.eps}
%\caption{\label{xft} Predicted abundance ratios in the dust as function of [Fe/H] for Models M9d (upper panel), M310d (middle panel), M12d (lower panel)..}
\caption{\label{xft} Predicted abundance ratios in the dust as function of [Fe/H] for a galaxy of $3 \cdot 10^{10}M_{\odot}$ (Model M310d).}
\end{figure}

Finally, in Fig.~\ref{xft} we show the predicted abundance ratios in the dust
for the lowest (M9d) and highest (M12d) mass cases compared to an
intermediate mass one (M310d).
The high [Mg/Fe] and [Si/Fe] are easily explained by the fact that Mg, Si and Fe have similar
condensation efficiencies, therefore the abundance ratios in the dust
mirror those in the gas. On the other hand O - the dominant
metal in the gas - is underabundant with respect to Fe in the dust.
In passing, we note that the trend in the dust [Mg/Fe], [Si/Fe] and [O/Fe] ratios versus [Fe/H] does not
significantly with the galaxy mass. 
We stress that these are actual predictions of the models which await observational
confirmation and, if so, will help in understanding the abundance pattern
in high redshift galaxies.

%\begin{figure}
%\caption{\label{xft} Predicted abundance ratios in the dust as function of [Fe/H] for a galaxy of $3 \cdot 10^{10}M_{\odot}$ (Model M310d).}
%\end{figure}

%\begin{figure}
%\includegraphics[width=9cm]{x_fe_Fe_h_in_dust_M11d.eps}
%\caption{\label{xft} Predicted abundance ratios in the dust as function of [Fe/H] for a galaxy of $3 \cdot 10^{10}M_{\odot}$ (Model M310d).}
%\end{figure}

%\begin{figure}
%\caption{\label{xft} Predicted abundance ratios in the dust as function of [Fe/H] for Model M310d.}
%\end{figure}

\section{The high mass case: comparison to a QSO host}

\begin{figure}
\includegraphics[width=9cm]{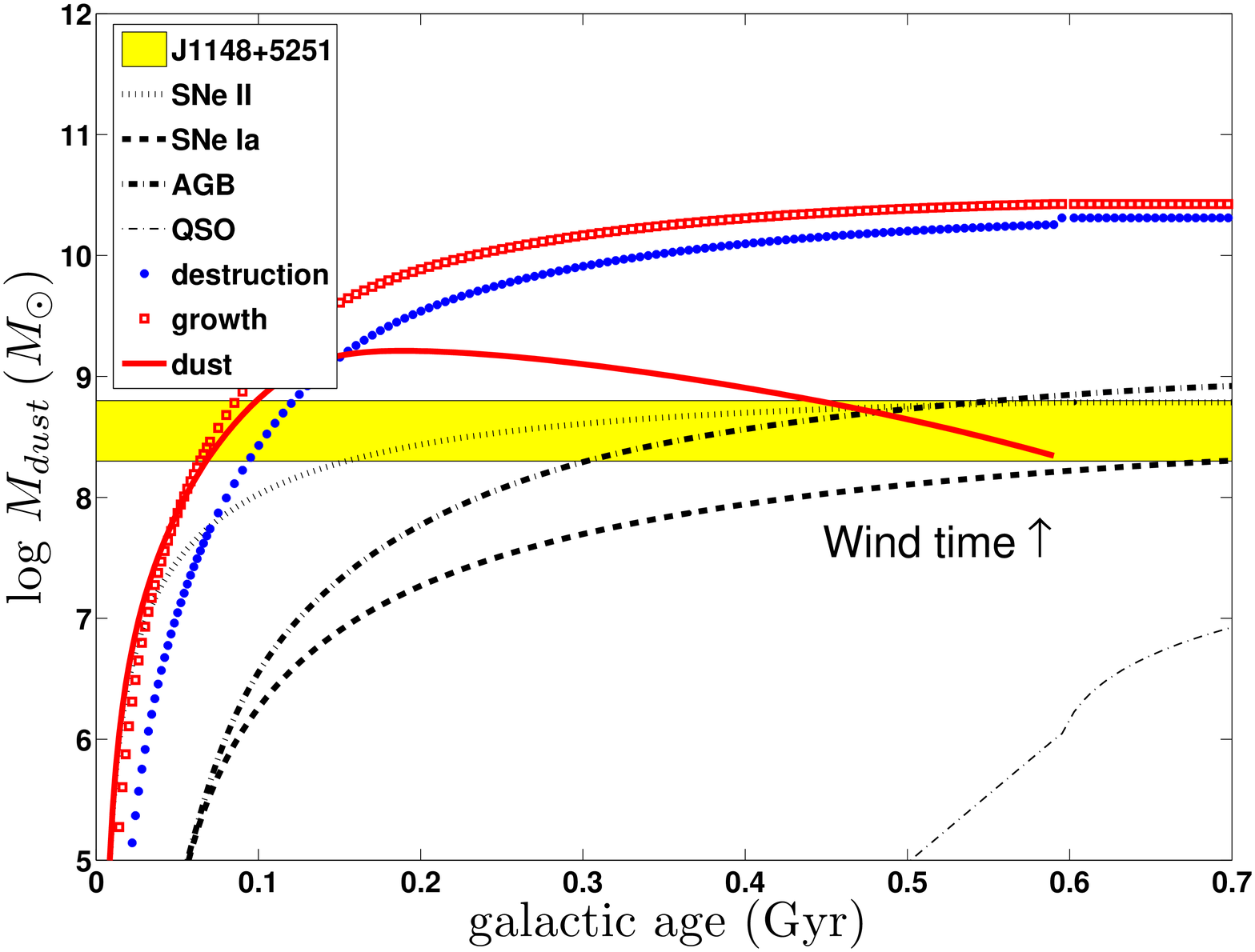}
\includegraphics[width=9cm]{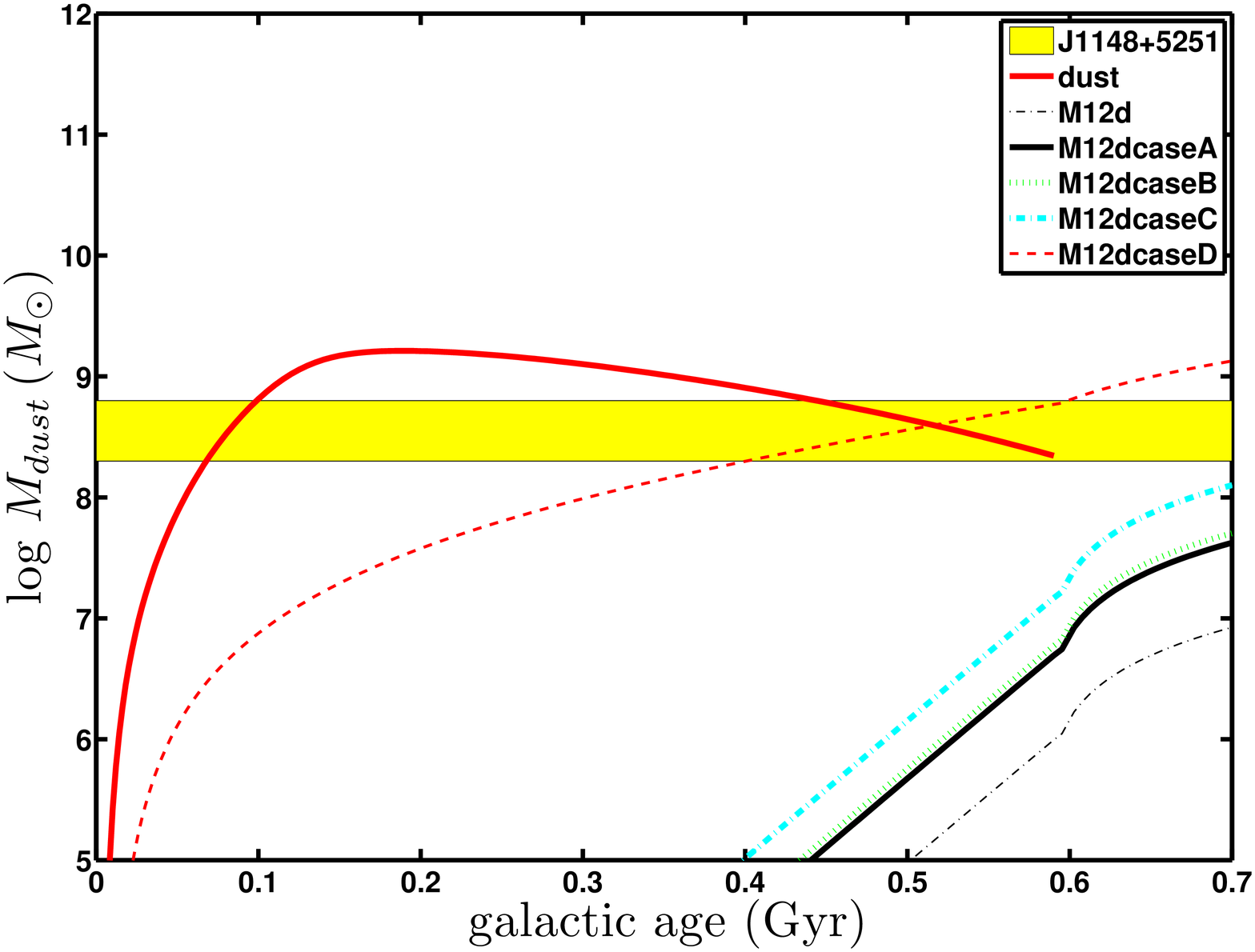}
\caption{\label{dustmass}  \emph{Upper panel}: Dust mass (Red solid line) as function of galactic age. The cumulative
contribution to the dust mass by SNIa, SNII, AGB and QSO (fiducial case)) as well as the integral of the the
dust growth and destruction are shown. The observed values are given by the shaded area. The arrow
shows the time at which we predict the conditions for the galactic wind to occur in most of the galaxy.
The model refers to a galaxy of $10^{12}M_{\odot}$ (model M12d).
\emph{Lower panel:} as above, in this case only the total dust and the QSO contribution are plotted. The QSO
fiducial dust production is compared to a case in which we assume the metallicity of the dust
forming regions 0.8 dex higher than the average (Case A), and to two cases in which
the relation between BH mass and stellar mass is ``super-Magorrian'' (Cases B and C, see text).
Finally, we present a model (Case D) in which the QSO is fine-tuned to be the only source
of dust.} % and a model which is based on the Calura et al. (2008) prescriptions.}
\end{figure}

\begin{figure}
\includegraphics[width=9cm]{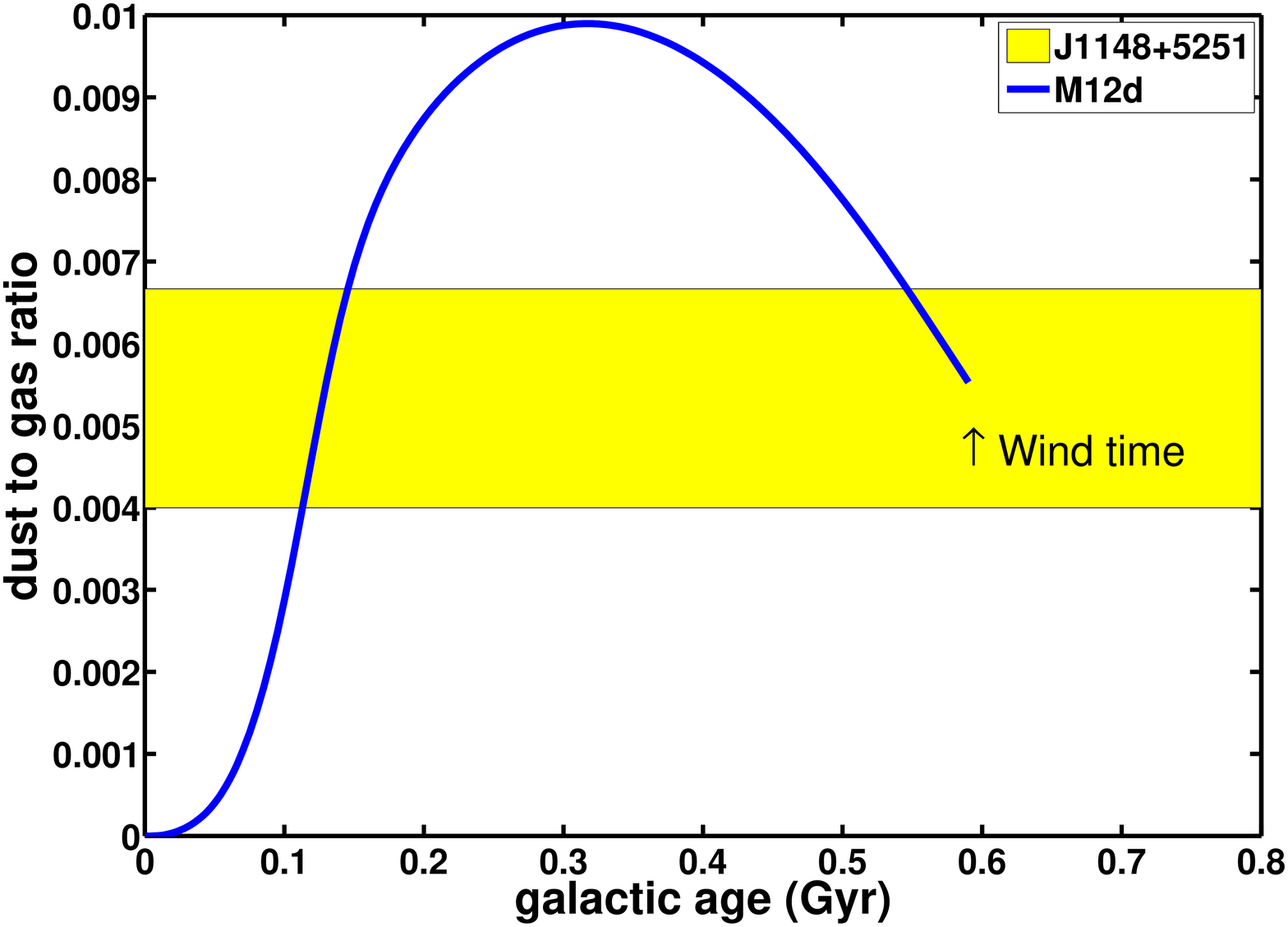}
\caption{The predicted dust to gas mass ratio as a function of galactic age for model M12d (continuous line) compared to the observational range of data (shaded area). \label{dg}}
\end{figure}

In this section we apply our model to the challenging case of the \qj \,  host galaxy.
{ In the first place, we note that our model predicts a stellar mass of $\sim 10^{11}M_{\odot}$
and a gas mass of $\sim 10^{10}M_{\odot}$ within 2.5 kpc (i.e. 1/4 $R_{eff}$)
in agreement with the observations within the same aperture (e.g. Wang et al., 2010).
The predicted BH mass at the onset of the galactic wind is $\sim 2\cdot 10^{9}M_{\odot}$,
as required by observations (e.g. Barth et al., 2003).}
In the upper panel of Fig.~\ref{dustmass} we show the predicted dust mass as a function
of galactic age  and compare it with the observed dust mass (shaded area). The model,
which  has been calibrated on the LBGs and present-day properties of ellipticals,
does reproduce the necessary amount of dust mass at $t\sim t_{gw}=0.6$ Gyr.
The fact that the dust content is even higher at earlier times is perhaps
suggestive of the evolution from a relatively long period of dust enshrouded star formation 
to a phase in which stellar + AGN driven winds make the galaxy visible and
the QSO shine.
Similar conclusions can be derived by using the predicted dust-to-gas ratio (Fig.~\ref{dg}).

\begin{figure}
\includegraphics[width=9cm]{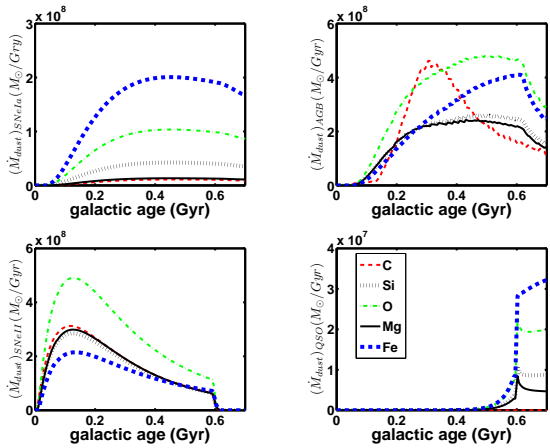}
\caption{\label{d_p_r}  Dust production rate of SNe Ia (left upper panel), SNe II(left lower panel), AGB stars (right upper panel), QSO(right lower panel) as functions of galactic age for different elements as indicated in the figure, for model M12d. }
\end{figure}

We now turn our attention to the details of the dust production.
In Fig.~\ref{dustmass} (upper panel) we also show the
cumulative mass of dust produced by each single channel (including the dust
growth) as well as the total mass of dust destroyed.
As it happens in the solar vicinity (Calura et al., 2008), dust growth
and destruction dominate at times larger than 0.1 Gyr. The QSO
production seems to be relatively unimportant. However, we must stress
that while SNII production stops and the dust growth is strongly suppressed after the galactic wind,
the QSO dust production rate should be at its maximum, as shown in Fig.~\ref{d_p_r}, at the time of the galactic wind.
It is also important to stress that the QSO dust should  be confined to the very central
regions of the galaxy, whereas the dust ejected by SNe Ia and AGBs will be more diffuse,
therefore the QSO is likely to be the dominant source right after the galactic
wind in the inner regions of the galaxy.

Although similarly high dust masses have been observed in other high redshift QSOs (Beelen et al., 2006),
\qj is interesting because at its redshift in the favored cosmology, the star formation, BH growth and dust 
production can be reproduced only assuming a very fast evolution. 
Several other authors have attempted to reproduce the observed dust mass: some
(e.g. Dwek et al., 2007a) claim that the host galaxy is too young to have had any
AGB contribution to the dust, whereas others (e.g. Valiante et al., 2009a) claim
that AGB stars are actually a viable channel. Moreover,
the observed dust mass might be consistent with QSO production (see Maiolino et al, 2006).
However, it is rather difficult to make detailed comparison between different models because
of their different assumptions. 

{ For instance, Valiante et al. (2009a) do not consider dust growth, have
a less efficient destruction (since they do not take into account SNe Ia),
but adopt much smaller dust depletion efficiencies.
Dwek et al. (2007a) assumed that, owing to the young
($<$ 0.5 Gyr) inferred age of the host galaxy, only SNII contribute to the dust production
and neglected the role of AGB. According to stellar evolution theory, instead, 
intermediate mass stars evolve out of the main sequence after just $\sim$ 30 Myr.
Therefore, we argue that Dwek (2007)'s assumption is too simplistic (see also Valiante et al., 2009a)
and a more detailed models of chemical
evolution as the one used in this paper is required to assess the AGB contribution}.
{ On the other hand, our fiducial model needs only 0.6 Gyr to reproduce dust, stellar, gas and QSO
masses at the redshift (6.4) at which we observe \qj . Such a time lapse
 would imply a (maximum) formation redshift of 13. This is the value that we adopt for
our model, so that redshift 6.4 corresponds to $t_{gw}$.
We note that the predicted dust mass for
our fiducial model crosses the region of the observed values in Fig.~\ref{dustmass}
also at earlier times (0.05 Gyr) which would imply - in the current favoured
cosmology - a (minimum) formation redshift of 6.8. Finally, according to Fig. 2, the 
predicted SFR crosses the observed value at $\sim$0.4 Gyr, hence implying a
formation redshift of $\sim$10.}

{ In a sense, our predictions give lower limits to the QSO contribution, which
can be higher if either the BH ``seed'' is larger and pre-existing to the galaxy
or the growth is super-Eddington in the very early stages and quieter
near the time of the galactic wind. Such a situation is depicted in the lower
panel of Fig.~\ref{dustmass}, where the fiducial QSO production (dot-dashed line) adopted here is
compared to other cases to show the sensitivity to our different assumptions.
The main properties are summarized in Table~\ref{t1}.
Case A (solid line) shows the case in which we consider the metallicity of the region where
the QSO dust forms 0.8 dex higher than the mean metallicity predicted by our model.
This is because BLRs exhibit metallicities up to 10 times the solar value, whereas
our model predicts the average gas metallicity of the central zone to be twice
than solar. Such a change in metallicity increases the QSO dust mass by about one
order of magnitude. A similar change can be achieved if we take into
account the fact that the BH-to-stellar mass ratio is a factor of 2 higher at
high redshift with respect to the local one (Case B, dotted line, e.g. Walter et al., 2004,
Maiolino et al., 2007, Lamastra et al., 2010, Merloni et al., 2010, Wang et al. 2010, but see Willott et al., 2010). To take
into account the uncertainty in the BH-to-stellar mass ratio at high redshift, we 
make an additional model in which this ratio is a factor of 4 higher than the local value (Case C, light
dot-dashed line). It is worth noting that in both Case B and C the final predicted $M_{BH}$ is still
consistent with observations, within the uncertainties.}
{ Finally, Case D is a model in which we switch off the dust production
from stars and assume that all the dust seen at $t\sim t_{gw}$ has been produced by the QSO.
As it can be inferred from comparing the fiducial model to cases A, B, C and D,
even if the QSO is the only source of dust, there will still be degeneracy
between the accretion timescale, the initial BH seed mass and the metallicity
of the clouds where the dust forms. The purpose of model M12caseD is, therefore,
to give just an example and show that a QSO alone might produce enough dust
if, for instance, we start from a quite high seed mass ($10^8 M_{\odot}$),
while the accretion timescale (0.49 Gyr) is 10 times longer than the adopted one in the fiducial case.}
Such a situation mirrors the state of the art: the channels for dust production
are many, and their relative role cannot be constrained by using only the
observed dust mass.
%The predicted dust production rates for each element for each channel are shown in 
%Fig.~\ref{d_p_r} for a galaxy of $10^{12}M_{\odot}$.

\subsection{The QSO abundance ratios}

D'Odorico et al. (2004) determined C, N and $\alpha$-element relative abundances in the gas surrounding six  QSOs (from the NLR) at an average redshift of $z \sim$  2.4, by studying six narrow associated absorption systems in Ultraviolet Visual Echelle Spectrograph (UVES) high-resolution spectra. They found five systems with a metallicity (measured by C/H) consistent with or above the solar value. They also found a possible correlation of [N/C] with [C/H] and an anticorrelation of [Si/C] vs. [C/H].
They suggested that to explain such abundance ratios the chemical enrichment in the host galaxy had to be very fast, not longer than 1 Gyr. This conclusion is in agreement with a previous work by Matteucci \& Padovani (1993) who reproduced the abundaces measured in the BLR of QSOs, in particular the oversolar Fe abundance and solar O abundance by means of a model of a large elliptical suffering a strong burst of star formation. In particular, the high [Fe/Mg] ratio claimed in some objects was interpreted as the value in the gas around the QSO after star formation has stopped. In fact, in such a situation, $\alpha$-elements are no more produced whereas Fe and N, formed in low and intermediate mass stars, are continuously produced, thus creating a situation where N and Fe are oversolar and the $\alpha$ elements are solar. 
Our model predicts oversolar [Mg/Fe] for the dominant stellar population and undersolar [Mg/Fe] in the gas after the galactic wind, in particular the average ratios after the galactic wind in the gas are [Mg/Fe]$\sim$-0.2 dex.
That explains also the almost constant metal content in QSOs at all redshifts and consequently the similar metallicities found in BLR at high redshift and in NLR at lower redshift (e.g. Nagao et al., 2006a, 2006b, Juarez et al., 2009): most of the evolution in these objects has occurred at very ealy epochs where SNe II produced most of the metals. After star formation stopped due to either galactic winds (Matteucci \& Padovani 1993) or AGN feedback (Romano et al. 2002), the SNe Ia produced the bulk of Fe and low mass stars the bulk of N which kept increasing for a while and then settled on a  rather constant value due to the decrease of Fe an N producers with time (no star formation takes place).
In Fig. 17 we show our predictions (solid line) for [N/C] and [Si/C] versus D'Odorico et al.'s data. There is good agreement with the observational trends. 
{ A marginal agreement (within $3\sigma$) with the} high [C/H] observed in  QSOs (exceeding $\sim +1.4$ dex) can be obtained only by adopting the Maeder (1992) yields with mass loss from massive stars (dotted lines), { which allow the model
to reach [C/H] values slightly larger than 0.8 dex}. With normal yields (e.g. Woosley \& Weaver 1995) we would no go beyond [C/H]=+0.5 dex.

\begin{figure}
\includegraphics[width=9cm]{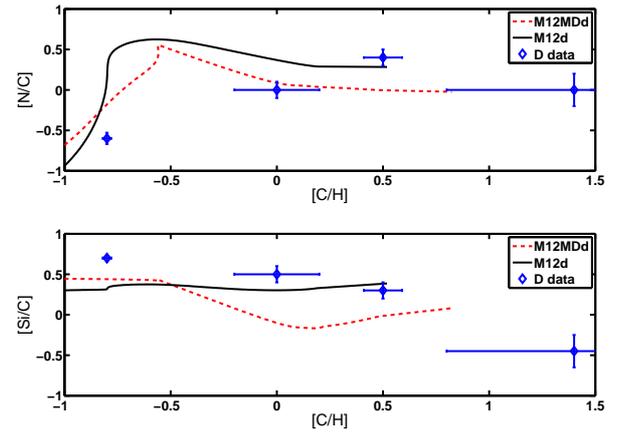}
\caption{\label{si_n_c}  Predicted and observed [N/C] and [Si/C] ratios versus [C/H]. The model refer to an elliptical of 
$10^{12}M_{\odot}$. In particular, the solid line refers to the fiducial case (model M12d), whereas the dotted
line gives the predictions in which the metallicity dependent yield from Woosley \& Weaver (1995) + Maeder (1992) are
adopted (model M12MDd). Data: NLR by D'Odorico et al. (2004).    }
\end{figure}

This result confirms that of Cescutti \& al. (2009) who found that the Galactic bulge high C abundances can be reproduced only by adopting yields from massive stars with mass loss strongly dependent on metallicity.

\section{Conclusions}

In this paper we have studied the evolution of several chemical elements (C, N, O, Si, Mg, Fe) both in the gas and in the dust in elliptical galaxies of different masses (from $10^{9}M_{\odot}$ to $10^{12}M_{\odot}$). Our aim was to study the evolution of gas and dust at high redshift and the effect of different environments. We assumed that dust is produced in: i) SNe II, ii) SNe Ia, iii) AGB stars and iv) QSOs. We took into account also dust destruction and accretion. These models will help up-coming observations of distant objects in quantifying their morphologies, past SFR
and mass by means of abundances and abundance ratios, as well as the role of the dust in the measurement of the above quantities.
In particular, our general predictions are that:
\begin{itemize}
\item The downsizing implies that at { [Fe/H]$<-1$} the most massive objects have lower [N/Fe] ratios than
the lower mass ones, irrespective of the presence of dust. { The same is true for the [N/O] ratios in a broader
range of metallicities.}
\item A dust mass-stellar mass relation exists, with more massive galaxies attaining a higher dust content at earlier time. 
Thererefore the expected evolution is similar to the observed high redshift evolution of the mass-metallicity relation.
\end{itemize}
These predictions await confirmation from future observations.
In order to test the model we compared our results for the lower mass galaxies with LBG galaxies and for the highest mass limit with QSOs. 
In particular, we compared our model for LBGs also with the specific galaxy \lbg. Unfortunately, the observed abundance ratios from different authors show very different values. Our predicted [N/Fe] ratio assuming no primary production of N from massive stars and with dust reproduces very well tha data of Pettini \& al. (2002). On the other hand, the model assuming dust and primary N production from massive stars predicts the [N/Fe] ratio in very good agreement with the data of Teplitz et al. (2000).  Therefore, no conclusions can be drawn on this point.

As far as these specific class of objects are concerned, our findings can be summarized as follows:
\begin{itemize}

\item LBG galaxies are likely to be young ellipticals of intermediate mass($10^{10}-3\cdot 10^{10} M_{\odot}$) experiencing moderate SFR and galactic winds. Our elliptical model for $3\cdot 10^{10} M_{\odot}$ well reproduces the [O/H] abundance as a function of redshift for the LBG galaxies. Our model implies a redshift of formation for these galaxies in the $z=2.3-4$ range.

\item The predicted [Mg/Fe] and [Si/Fe] ratios have been compared with data for \lbg \, by Pettini \& al. (2002) 
and the theoretical values are lower than the observed ones, perhaps indicating that too low a fraction of Fe is predicted to be in dust. 
%The dust evolution is a complicate issue since it depends on the rate of dust production, 
%destruction and accretion and many uncertainties are still present in these processes.

\item From the comparison of our model results for a massive elliptical 
($M=10^{12}M_{\odot}$) with one of the most distant QSO ever observed J1148+5251, we derived a redshift of formation  for this object $z> 7$ (and likely larger than 10).

\item The predicted total amount of dust is in good
agreement with the dust seen in this QSO. We tested also the hypothesis that the QSO itself produced dust 
but this production appears negligible compared to that from stellar sources, unless one focuses
on the very central regions at times very close to the galactic wind onset.
Future observations can provide better constraints to the QSO dust model.

\item We also compared our model results for QSO hosts with data from NLR in QSO hosts and we found a good agreement for [N/C] versus [C/H] and [Si/C] versus [C/H].
The very high C abundance observed in these QSOs can be explained only by assuming yields with mass loss from massive stars with a strong dependence on metallicity, as those of Maeder (1992).

\end{itemize}

\section{Acknowledgments}
We thank the anonymous referee for comments which improved the quality of the presentation.
AP acknowledges useful discussions with G.Cescutti, M.Elvis, A.Shapley and R.Valiante and
thanks K. Kornei for a careful reading of the paper.
AP receives support from Italian Space Agency
through contract ASI-INAF I/016/07/0 %and  from M.Rich's (XXX) grant.
F.M acknowledges financial support from PRIN2007 MIUR (Italian Ministry of Scientific Research) 
project Prot. no. 2007JJC53X-001.

\end{document}